\documentclass[aps, prl, reprint, superscriptaddress, final]{revtex4-1}
\usepackage[utf8]{inputenc}
\usepackage{amsmath}
\usepackage{amsfonts}
\usepackage{amssymb}
\usepackage{graphicx}
\usepackage[obeyDraft]{todonotes}

\begin{document}
\title{Electron nuclear double resonance with donor-bound excitons in silicon}

\author{David P. Franke}
\email{david.franke@wsi.tum.de}
\author{Michael Szech}
\author{Florian M. Hrubesch}
\affiliation{Walter Schottky Institut and Physik-Department, Technische Universit\"at M\"unchen, 
	85748 Garching, Germany}

\author{Helge Riemann}
\author{Nikolai V. Abrosimov}
\affiliation{Leibnitz-Institut für Kristallz\"uchtung, 12489 Berlin, Germany}

\author{Peter Becker}
\affiliation{PTB Braunschweig, 38116 Braunschweig, Germany}

\author{Hans-Joachim Pohl}
\affiliation{VITCON Projectconsult GmbH, 07743 Jena, Germany}

\author{Kohei M. Itoh}
\affiliation{School of Fundamental Science and Technology, Keio University, Yokohama 223-8522, Japan}

\author{Michael L. W. Thewalt}
\affiliation{Simon Fraser University, Burnaby, British Columbia V5A 1S6, Canada}

\author{Martin S. Brandt}
\affiliation{Walter Schottky Institut and Physik-Department, Technische Universit\"at M\"unchen,	85748 Garching, Germany}

\begin{abstract}
	We present Auger-electron-detected magnetic resonance (AEDMR) experiments on phosphorus donors in silicon, where the selective optical generation of donor-bound excitons is used for the electrical detection of the electron spin state. Because of the long dephasing times of the electron spins in isotopically purified $^{28}$Si, weak microwave fields are sufficient, which allow to realize broadband AEDMR in a commercial ESR resonator. Implementing Auger-electron-detected ENDOR, we further demonstrate the optically-assisted control of the nuclear spin under conditions where the hyperfine splitting is not resolved in the optical spectrum. Compared to previous studies, this significantly relaxes the requirements on the sample and the experimental setup, e.g.~with respect to strain, isotopic purity and temperature. We show AEDMR of phosphorus donors in silicon with natural isotope composition, and discuss the feasibility of ENDOR measurements also in this system.
\end{abstract}

\maketitle
Spin-to-charge conversion in electrically detected magnetic resonance provides a very sensitive way of measuring the spins of donors in silicon \cite{stegner_electrical_2006, mccamey_electrically_2006, mccamey_electronic_2010, steger_quantum_2012}, enabling the detection of single spins \cite{morello_single-shot_2010, pla_high-fidelity_2013}, as well as spin resonance experiments at low or zero external magnetic field \cite{morishita_electrical_2009, franke_spin-dependent_2014-1, dreher_pulsed_2015}.
In the context of quantum computation, where the electron and nuclear spins of donors in silicon are interesting because of their extremely long coherence times \cite{tyryshkin_electron_2003,tyryshkin_electron_2012,steger_quantum_2012,saeedi_room-temperature_2013, itoh_isotope_2014}, electrical detection, as well as electrical control \cite{bradbury_stark_2006, lo_stark_2014, wolfowicz_conditional_2014,laucht_electrically_2015}, could facilitate an integration of quantum bits with current semiconductor technology.
In the case of electrical detection based on spin-dependent recombination, the time scales which can be addressed can be limited by the spin lifetime of the particular readout partner \cite{hoehne_time_2013, franke_spin-dependent_2014-1} which in turn also limits the nuclear spin coherence time of neutral donors \cite{dreher_nuclear_2012, hoehne_submillisecond_2015}.
The spin-to-charge conversion based on the creation of donor-bound excitons (DBE), on the other hand, does not have this effect \cite{steger_quantum_2012}.
DBE complexes can be formed by resonant infrared laser excitation and almost immediately decay in an Auger process generating nonequilibrium charge carriers in the conduction band, allowing the detection of the optical transition as a photocurrent. In particular in isotopically purified $^{28}$Si, these infrared transitions can be remarkably sharp and the Zeeman interaction of the donor and the DBE complex with magnetic fields can be resolved in the optical spectra \cite{karaiskaj_photoluminescence_2001, yang_optical_2006}. This spin-selective excitation can be used as an electrical detection mechanism for the electron spin of the donor \cite{yang_optical_2006}, called Auger-electron-detected magnetic resonance (AEDMR) \cite{steger_quantum_2012}. In samples with high purity and at low temperatures (typically $T<5$ K), it is even possible to optically resolve the hyperfine splitting %$A=117.5$ MHz \cite{feher_electron_1959-1}
\cite{yang_optical_2006, yang_simultaneous_2009} enabling the optical \cite{steger_optically-detected_2011, salvail_optically_2015} and electrical \cite{steger_quantum_2012} detection of the nuclear spin state.

In silicon with natural isotope composition ($^\mathrm{nat}$Si), the line broadening connected to random positions of the different isotopes in the host crystal renders this selective excitation of donors in a certain nuclear spin state impossible \cite{kaminskii_luminescence_1980, thewalt_direct_2007}. While electron spin selectivity via DBE in $^\mathrm{nat}$Si is obtained at high magnetic fields and allows for a fast and high polarization of the electron spins, an optical polarization of the nuclear spins can only be reached on a timescale of minutes to hours, based on Overhauser relaxation \cite{overhauser_polarization_1953,feher_electron_1959,dluhy_switchable_2015}. A similar situation arises for $^{28}$Si when the DBE lines are broadened by even weak inhomogeneous strains. Therefore in these experiments the photoconductivity is usually monitored in a contactless capacitive fashion by placing the sample between two metal plates and measuring the impedance of this assembly. Only recently, AEDMR of the electron spin of $^{31}$P donors was realized in a $^{28}$Si sample equipped with evaporated Al contacts \cite{lo_hybrid_2015}. While this experimental realization allows for a higher sensitivity compared to a capacitive detection of the sample's conductivity, the strain induced by the contact structure appears to significantly shift and broaden the DBE transitions, inhibiting the separation of the nuclear spin states in that study.
We address this issue by implementing Auger-electron-detected electron nuclear double resonance (ENDOR), enabling us to detect the nuclear spin state of $^{31}$P donors in experimental conditions such as higher temperatures or the presence of strain, where the hyperfine interaction is not optically resolved. We further show the feasibility of AEDMR in $^\mathrm{nat}$Si at magnetic fields corresponding to X-band frequencies, demonstrating that our ENDOR approach will also be beneficial in such samples.

\begin{figure}
	\centering
	\includegraphics[width=\linewidth]{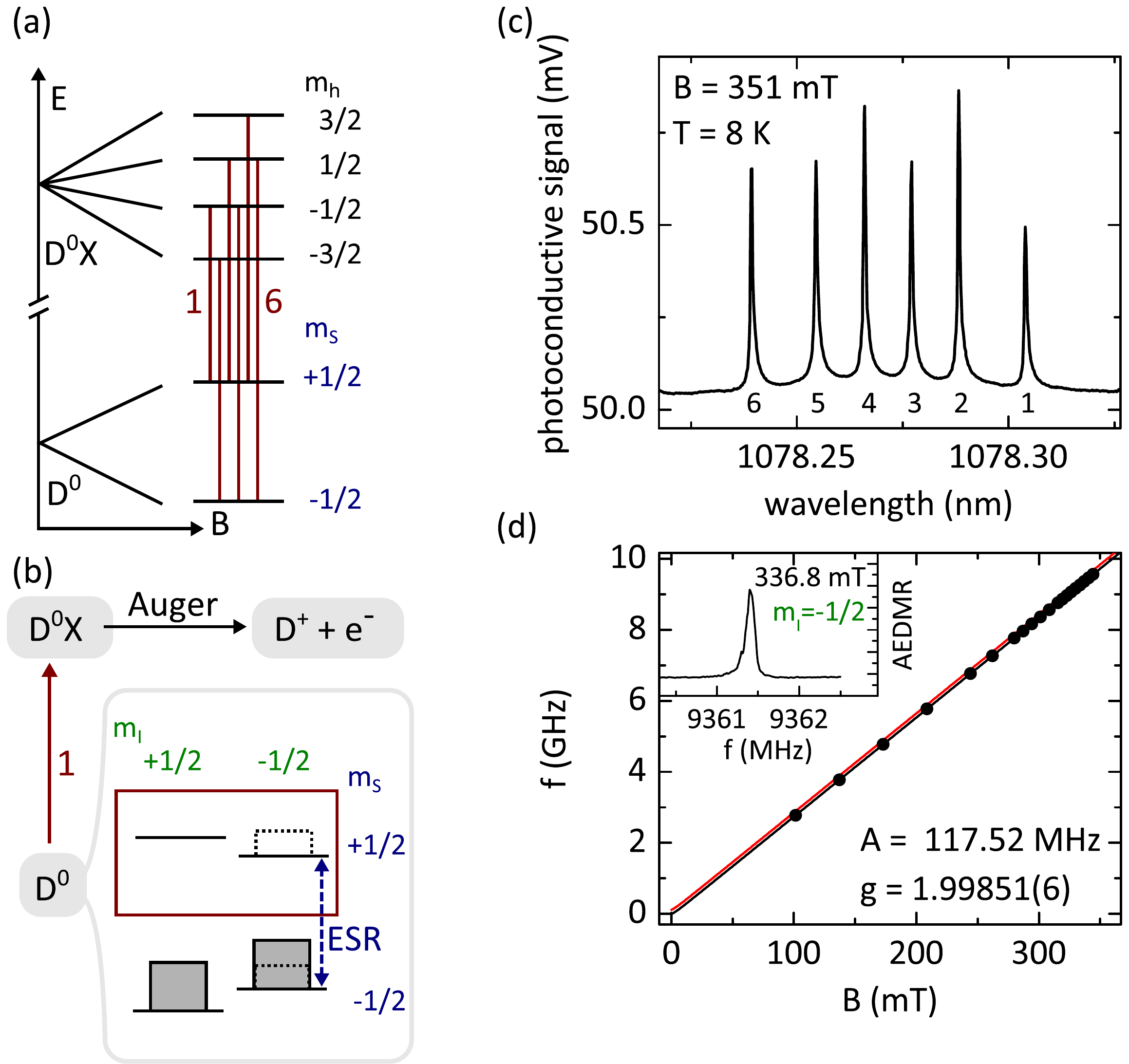}
	\caption{
		(a) Level scheme showing the donor-bound exciton transitions. (b) Level population scheme demonstrating the principle of AEDMR. For details see text. (c) Photoconductive spectrum at $B=351$ mT showing the six electron-spin-selective transitions. (d) Frequency of the $m_I=-1/2$ resonance as a function of the magnetic field $B$. Solid black and red lines represent calculated $m_I=-1/2$ and $m_I=+1/2$ peak positions, respectively, for $g=1.99851$ and a hyperfine interaction $A=117.52$ MHz.}
	\label{fig:scheme}
\end{figure}

A schematic representation of the DBE transitions (not considering the donor nuclear spin) is shown in Fig.~\ref{fig:scheme}(a). The neutral donor states are labeled D$^ 0$ and split up in two levels with electron spin projections $m_S=\pm1/2$ due to the Zeeman interaction with an external magnetic field $B$. The donor-bound exciton states, labeled D$^0$X, split according to the Zeeman interaction of the hole with spin 3/2 and spin projections $m_h=-3/2\dots 3/2$, since the two electrons in the three-particle DBE complex form a spin singlet. This leads to a total of six dipole-allowed optical transitions indicated by the red lines labeled 1 through 6 \cite{yang_optical_2006}.
The selective excitation of one of these lines allows us to perform electrically detected ESR experiments, as schematically shown in Fig.~\ref{fig:scheme}(b), where the laser is assumed to excite DBE line 1. The ionization of donors in the $m_S=+1/2$ state and the subsequent capture of an electron with random spin state lead to an accumulation of donors with $m_S=-1/2$ and hence a very large steady-state polarization. The level populations of the neutral donors D$^0$ with electron and nuclear spin projections $m_S$ and $m_I$, resp., are shown by the gray boxes in Fig.~\ref{fig:scheme}(b). Since the ESR transitions, in contrast to the DBE transitions in our experiments, are nuclear-spin-selective, all four donor levels have to be considered when discussing the principle of the AEDMR experiments.
If one of the two ESR transitions ($\Delta m_S=\pm1$, $\Delta m_I=0$) is saturated by microwave (mw) irradiation (dashed blue arrow), one of the $m_S=+1/2$ states gets repopulated (dashed boxes). Under continuous excitation of one of the DBE lines (here: DBE line 1), the ESR saturation enhances the formation of DBEs and, subsequently, the ejection of charge carriers into the conduction band. It is easily seen that, while shown for DBE line 1 and $m_I=-1/2$ here, this process can realize an electrical detection of ESR for all six DBE lines and both ESR transitions.

The samples used in this work are an isotopically purified $^{28}$Si sample ($[^{28}\mathrm{Si}]=99.995 \%$) with a dopant concentration $[\mathrm{P}]=8\times 10^{14}~\mathrm{cm}^{-3}$ (size $\sim 15\times 3\times 1$ mm$^3$) \cite{steger_optically-detected_2011} and a $^\mathrm{nat}$Si sample cut from a commercial float zone wafer with $[\mathrm{P}]=5\times 10^{15}~\mathrm{cm}^{-3}$ (size $\sim 15\times 4\times 1.5$ mm$^3$). The experiments are performed in a Bruker flexline X-band resonator for pulsed ENDOR in a He-flow cryostat at a typical temperature $T=6$ K. Microwave pulses are defined by a digital pulse card generating square pulses that are mixed with continuous-wave (cw) microwave. These are then amplified by a traveling-wave-tube amplifier and attenuated, resulting in a typical mw power of $\sim 10$ W which in our system corresponds to a pulselength of $\sim 250$ ns for a $\pi$-pulse.
To avoid mechanical stress, the samples are mounted loosely between two gold-covered plates, the impedance of which is monitored at 476 kHz with a lock-in amplifier. The phase of the detection is chosen such that the signal-to-noise ratio of the photoconductivity measurement is optimal. The magnetic field was calibrated with an NMR Gaussmeter placed at the sample position, giving an estimated uncertainty of $\pm0.01$ mT. The NKT Photonics fiber laser provides wavelengths $\lambda$ between 1077.7 and 1078.5 nm, achieved by adjustment of the fiber temperature and an additional, fast tuning via the voltage applied to a piezo-electric crystal. A laser power of $15$ mW is used and the light is focused on the sample's thinner edge, the spot size on the sample is $\sim 1\times 5$ mm$^2$. Because of the very weak absorption \cite{thewalt_direct_2007}, we assume that the we probe the full depth of the samples with an estimated number of $10^{13}$ ($^{28}$Si sample) and $10^{14}$ ($^\mathrm{nat}$Si sample) phosphorus spins. For cw ESR measurements, the laser wavelength is tuned to one of the DBE lines and stabilized by a PI controller using the observed photoconductivity as feedback.

Figure \ref{fig:scheme}(c) exemplarily shows the measured photoconductive signal as a function of the laser wavelength and clearly shows the six DBE lines which are split by an external magnetic field $B=351$ mT.
Tuning the laser to DBE line 1 and sweeping the mw frequency $f$ at a fixed external magnetic field, we record AEDMR spectra. A typical AEDMR spectrum is shown in the inset of Fig.~\ref{fig:scheme}(d), where the photoconductive signal is shown as a function of the mw frequency for $B=336.8$ mT. Due to the long coherence times of the electron spin, only weak microwave powers are needed to saturate the ESR transition. Therefore, in cw experiments, we are able to use the microwave antenna of the X-band resonator as a broadband mw delivery system, allowing us to perform measurements over a large frequency and magnetic field range. The observed resonance positions are shown as circles in Fig.~\ref{fig:scheme}(d) and the theoretically expected positions for the low frequency ($m_I=-1/2$) peak are fit to the data (black line). With the hyperfine interaction constant $A=117.52$ MHz as determined by the ENDOR experiments below, the fit results in an electronic g-factor $g=1.99851(6)$, which is in very good agreement with previous measurements \cite{feher_electron_1959-1}.

\begin{figure}
	\centering 
	\includegraphics[width=\linewidth]{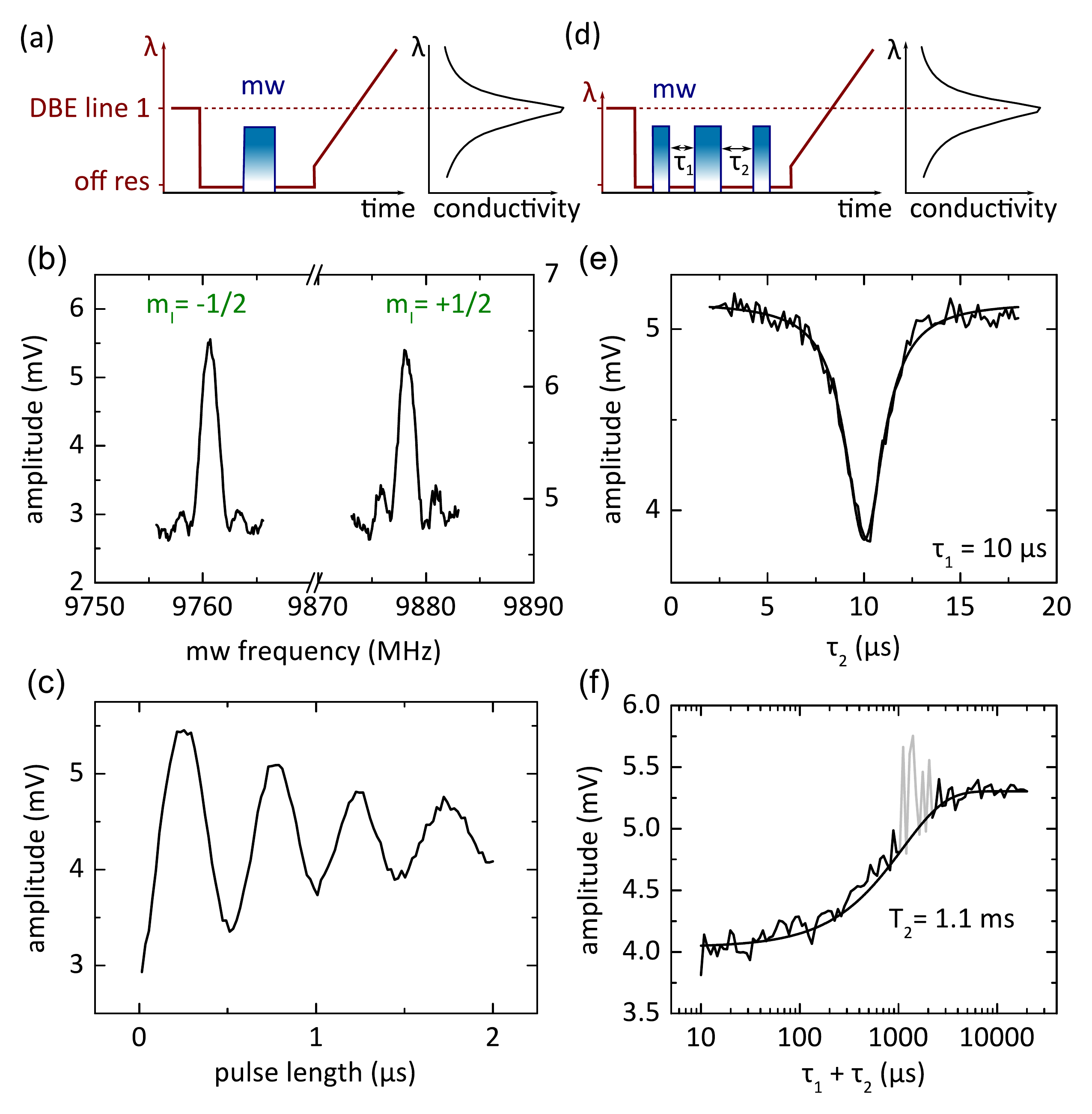}
	\caption{
		Auger-detected electron spin resonance experiments on $^{28}$Si. (a) Pulse sequence showing the laser tuning (red line) and an mw pulse (blue), as well a typical conductivity trace. (b) Resonance spectra and (c) Rabi oscillations of the $^{31}$P electron spin. (d) Pulse sequence used for electron spin echo experiments. (e) Echo for $\tau_1=10~\mu$s, (f) echo decay, revealing a coherence time $T_2=1.1$ ms. The data points shown in gray have been excluded from the exponential fit.}
	\label{fig:ADMR}
\end{figure}

To perform pulsed ESR measurements, we use the mw resonator, which was tuned to a low quality factor, around its resonance frequency and take advantage of the piezo-tuning of the laser to switch between on-resonance and off-resonance illumination. A typical pulse sequence is shown in Fig.~\ref{fig:ADMR}(a), where the laser tuning during the sequence is indicated by the red line, schematically showing the laser wavelength $\lambda$ on the vertical axis. After a long (typically 1s) polarization period (here DBE line 1), the laser is tuned off-resonance to avoid spurious ionization during the manipulation of the spin system via magnetic resonance. Then, one or more microwave pulses are applied. At the end of each sequence, the laser wavelength is swept over the DBE resonance line. A typical photoconductive trace as measured after the application of a resonant mw pulse is shown as a function of $\lambda$ at the right of the sequence. The amplitude of the observed line is determined by fitting the trace with a Lorentzian and is a measure for the population of the probed donor state (here $m_S=+1/2$). Note that the strong DBE signal allows us to perform all experiments single-shot, without additional averaging. Furthermore, this approach provides an easy and frequent calibration of the position of the DBE line to compensate for slow drifts in the laser wavelength or changes in the magnetic field.

We now apply mw pulses corresponding to a rotation of $\pi$ of the spin system, and again set the magnetic field to $B=351$ mT. ESR spectra are recorded by sweeping the mw frequency and clearly show the two hyperfine-split resonance peaks [Fig.~\ref{fig:ADMR}(b)]. As expected and analogous to the cw measurements, the photoconductivity is enhanced on resonance, where a repopulation of the $m_S=+1/2$ level is achieved (cf.~Fig.~\ref{fig:scheme}(b)). The lineshape is limited by the excitation bandwidth of the mw pulse (pulselength 500 ns), as indicated by the pattern reflecting the frequency distribution of the square pulse. By changing the length of the mw pulse, Rabi oscillations are recorded as shown in Fig.~\ref{fig:ADMR}(c), demonstrating that the coherent control of the electron spin can be detected. 
As can be deduced from the different amplitudes of the two hyperfine-split AEDMR lines, a nuclear spin polarization is created, most likely because of Overhauser electron spin relaxation via the hyperfine interaction \cite{overhauser_polarization_1953,feher_electron_1959}. It can reach up to 95\% after long periods of resonant illumination and can significantly hinder some of the measurements, such as AEDMR on both hyperfine-split lines. Therefore, resonant excitation of a different DBE line or above-bandgap illumination was sometimes used to reset the nuclear spin hyperpolarization between experiments \cite{hoehne_submillisecond_2015}.
Since it acts as an upper bound in the nuclear spin polarization process, we can estimate that the electron spin polarization in our experiments is close to 100\%. At the applied laser power, it is created within $\sim 1$ s, as determined by measuring the time constant of the transient current after application of resonant laser light.

For the detection of an electron spin echo, the single mw pulse is replaced by an echo sequence $\pi/2-\tau_1-\pi - \tau_2 - \pi/2$ which includes a final $\pi/2$ pulse \cite{breiland_optically_1973, huebl_spin_2008} to project the magnetization onto the $z$-axis where it can be detected by the AEDMR readout [Fig.~\ref{fig:ADMR}(d)]. The resulting spin echo is shown in Fig.~\ref{fig:ADMR}(e) for $\tau_1=10~\mu$s. We can measure the electron spin coherence time $T_2$ by recording the echo amplitude as a function of $\tau_1+\tau_2$ for $\tau_1=\tau_2$, which is shown in Fig.~\ref{fig:ADMR}(f), and find a $T_2=1.1$ ms, comparable to similar measurements on isotopically purified $^{28}$Si \cite{tyryshkin_electron_2012, lo_hybrid_2015}.

\begin{figure}
	\centering
	\includegraphics[width=\linewidth]{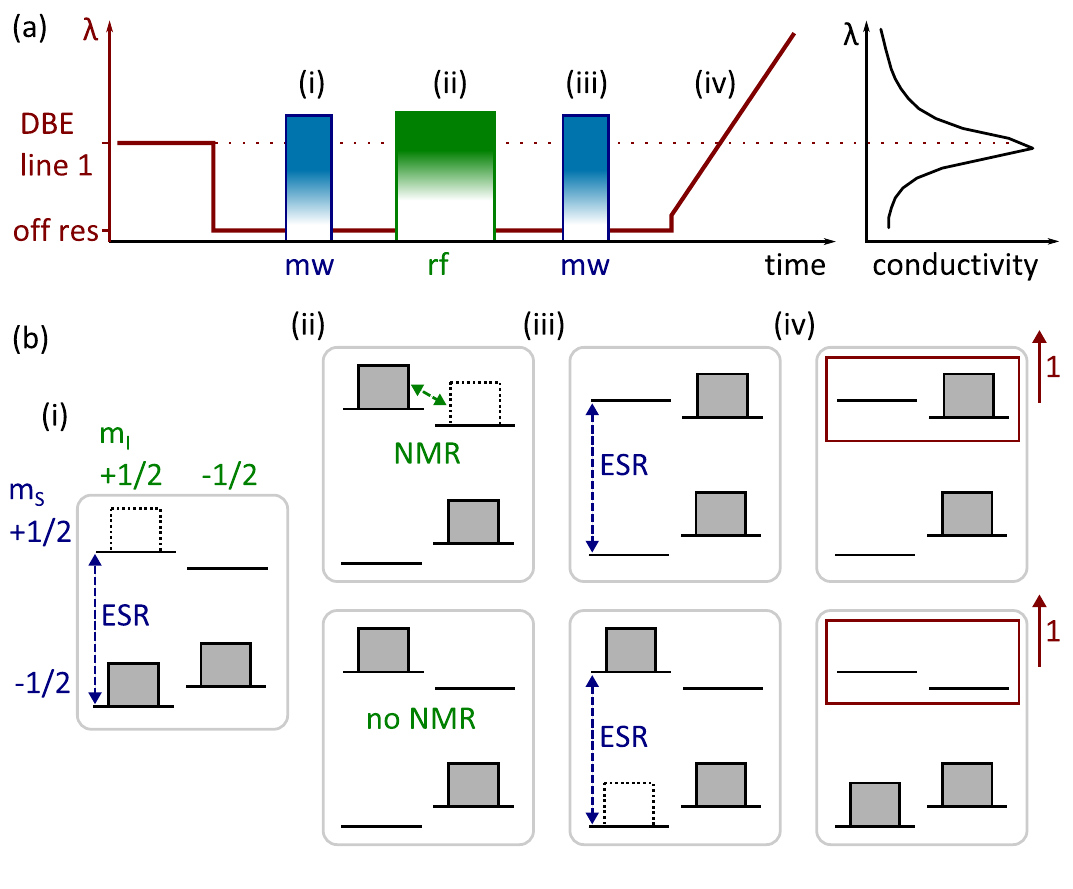}
	\caption{
		Schematic representation of the Auger-electron-detected ENDOR experiments. (a) Pulse sequence showing the laser tuning (red line), the mw (blue) and rf (green) pulses, as well a typical conductivity trace. (b) Level population diagram at points (i)-(iv) indicated in (a). Boxes represent the level populations before (solid boxes) and after (dashed boxes) the mw and rf transitions indicated by blue and green dashed arrows, respectively. The results of resonant (NMR) and off-resonant (no NMR) rf pulses are discussed in the top row and the bottom row, respectively.}
	\label{fig:ENDORscheme}
\end{figure}

For the detection and coherent manipulation of the nuclear spin, we combine mw and radio frequency (rf) pulses to implement Auger-electron-detected ENDOR (AEDENDOR). In silicon, ENDOR is frequently used in combination with conventional ESR detection \cite{feher_electron_1959-1, feher_donor_1959,hale_shallow_1969, tyryshkin_davies_2006, morton_solid-state_2008,george_electron_2010, morley_initialization_2010, simmons_entanglement_2011} and spin-dependent recombination \cite{stich_electrical_1996, hoehne_electrical_2011, dreher_nuclear_2012, hoehne_submillisecond_2015} to realize NMR experiments exploiting the higher sensitivity of electron spin detection.
We discuss our experimental approach with the help of Fig.~\ref{fig:ENDORscheme}, where the applied pulse sequence is shown in (a) together with level schemes illustrating the population of the donor spin states during the experiment in (b). At the beginning, the resonant laser excitation (DBE line 1) once more polarizes the system by ionization of the donors with  $m_S=+1/2$. As a result, only the $m_S=-1/2$ states are occupied as depicted by the gray boxes in Fig.~\ref{fig:ENDORscheme}(b)(i). After the laser is tuned off-resonance, a selective mw $\pi$ pulse swaps the populations of the two $m_I=+1/2$ states (dashed blue arrow).
Subsequently, an rf pulse is applied (ii) that is either resonant (NMR, top row) or off resonance (no NMR, bottom row) with the donor nuclear spin. In the case of NMR, a transition within the a subensemble (here: $m_S=+1/2$) is induced and the level populations are changed (green arrow). If the rf pulse is off-resonance, the populations remain unchanged.
Then, a second mw $\pi$ pulse is applied (dashed blue arrow in (iii)), which again swaps the populations of the two $m_I=+1/2$ states. The resulting configuration is different in the two cases discussed in Fig.~\ref{fig:ENDORscheme}(b). In the case of an off-resonant rf pulse (bottom row), the mw pulse swaps the populations back to their initial state. In the configuration after a resonant rf pulse (top row), however, the second mw pulse has no effect on the populations, since it is between two empty levels. Hence, one of the $m_S=+1/2$ levels remains occupied after the second mw pulse. At the end of the sequence, the laser wavelength is again swept and reveals the signal for DBE line 1, measuring the population of the $m_S=+1/2$ states (red frame in (iv)). Its magnitude is therefore enhanced in the case of a resonant rf pulse (top row). In principle, both NMR transitions can be measured using the same DBE line for polarization and detection of the spin ensemble. This becomes clear from Fig.~\ref{fig:ENDORscheme}(b)(ii). The application of an rf pulse resonant with the $m_S=-1/2$ subensemble populates the $m_S=-1/2$, $m_I=+1/2$ level, meaning that the final mw $\pi$-pulse is between two equally occupied levels. Similar to the case discussed above, this leads to a remaining occupation of one of the $m_S=+1/2$ levels and hence to a photoconductive signal.

\begin{figure}
	\centering
	\includegraphics[width=\linewidth]{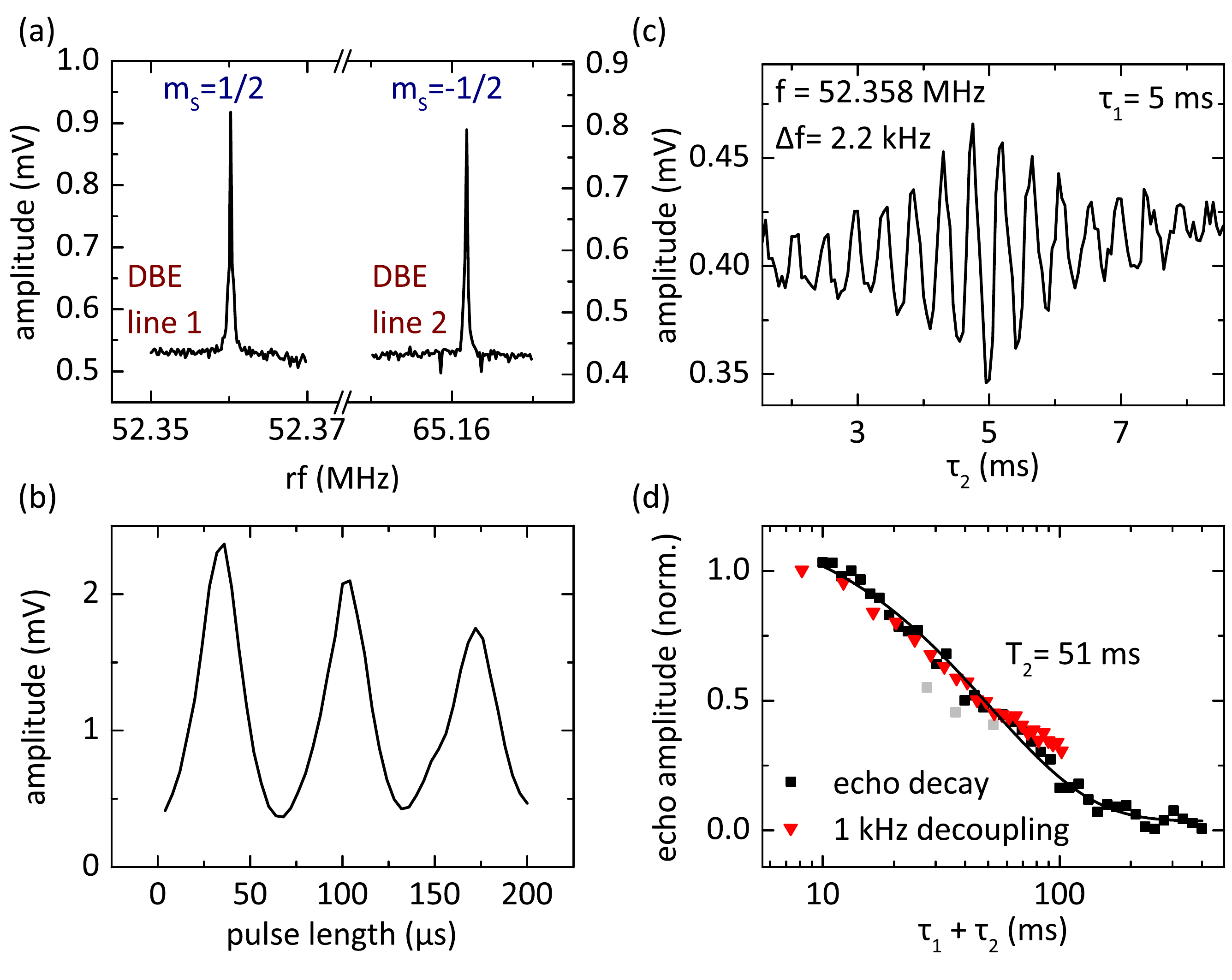}
	\caption{
		(a) Auger-electron-detected ENDOR spectrum showing the two $^{31}$P nuclear spin resonances. (b) Rabi oscillations of the nuclear spin. (c) Nuclear spin echo. The rf pulses are slightly off resonance ($\Delta f\approx 2.2$ kHz). (d) Nuclear spin echo decay, which is well described by an exponential decay (black line), revealing a coherence time $T_2=51$ ms, which is not significantly changed by the application of dynamical decoupling pulses (red triangles).}
	\label{fig:ENDORdata}
\end{figure}

For the experimental realization of the measuring scheme, we use the $m_I=+1/2$ ESR transition at $9760$ MHz and $B=351$ mT. In Fig.~\ref{fig:ENDORdata}(a), the resulting ENDOR signal is shown as a function of the frequency of the applied rf pulse. Two peaks are observed, corresponding to the resonances of the $m_S=+1/2$ and $-1/2$ subensembles. The line positions correspond to a hyperfine interaction of $A=117.52(2)$ MHz and a nuclear $g$-factor $g_n=2.259(2)$, in very good agreement with previous measurements \cite{feher_electron_1959-1, steger_optically-detected_2011}. As mentioned, both nuclear spin resonances can be detected with the excitation of the same DBE line. However, we have found that driving the NMR transition between the two levels that are not ionized by the laser leads to very long polarization lifetimes ($>1$ min) and significant broadening of the spectrum even when a $\pi/2$ reset pulse is added to the sequence. Therefore, we have used DBE line 2 which ionizes the $m_S=-1/2$ states for the detection of the resonance at $65.162$ MHz.

By changing the length of the applied rf pulse, we are able to record Rabi oscillations of the nuclear spin which are shown in Fig.~\ref{fig:ENDORdata}(b).
Replacing the single rf pulse by a three-pulse echo sequence equivalent to the electron spin echo sequence shown in Fig.~\ref{fig:ADMR}(c), a nuclear spin echo is recorded for $\tau_1=5$ ms [Fig.~\ref{fig:ENDORdata}(c)]. The echo displays an oscillation with $\Delta f\approx 2.2 $ kHz, revealing a slight offset of the applied radio frequency from the NMR transition. The width of the observed echo ($\sim 4$ ms) corresponds to a linewidth of about $250$ Hz, suggesting that the spectra in Fig.~\ref{fig:ENDORdata}(a) are also limited by the excitation bandwidth of the applied pulses (pulse length $3.5 $ ms)

\begin{figure}
	\centering
	\includegraphics[width=\linewidth]{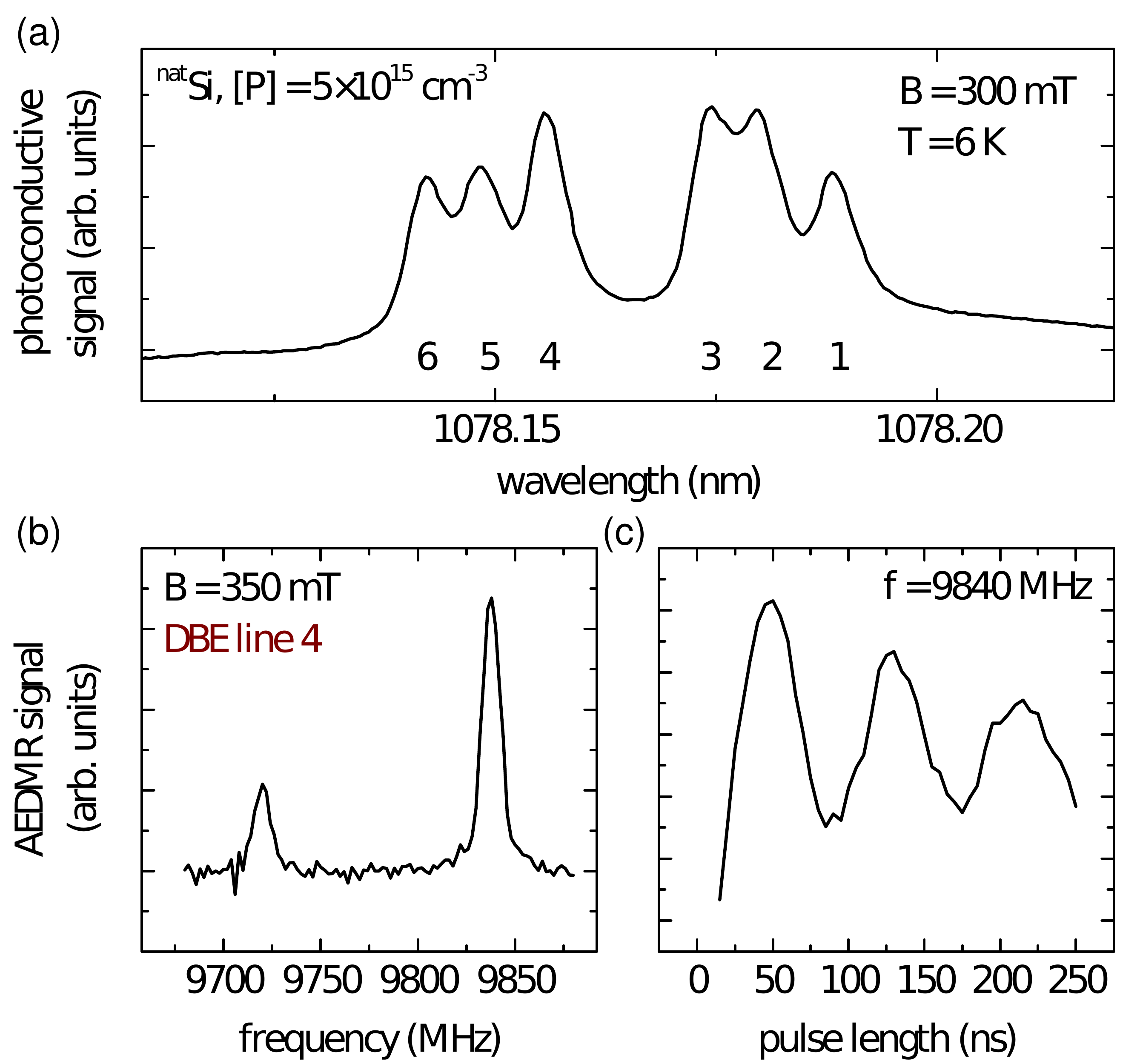}
	\caption{
		(a) Photoconductive spectrum of phosphorus-doped silicon with natural isotope composition at $B=350$ mT. (b) Pulsed AEDMR spectrum of the same sample. (c) Rabi oscillation of the phosphorus electron spin.}
	\label{fig:natSi}
\end{figure}

To determine the nuclear spin coherence time, we measure the echo amplitude as a function of $\tau_1+\tau_2$. The observed decay is well described by an exponential decay with a time constant $T_2=51$ ms. It is not significantly enhanced by the application of dynamical decoupling pulses, as shown by the red triangles in Fig.~\ref{fig:ENDORdata}(d), where refocusing pulses are applied at a frequency $f_{\pi}=1.01$ kHz in a Carr-Purcell sequence \cite{carr_effects_1954}. This suggests that the coherence time is limited by a process that is either frequency independent or cannot be refocused and is in agreement with the observation of an exponential decay in Fig.~\ref{fig:ENDORdata}(d) \cite{medford_scaling_2012}. While the electron spin lifetime $T_1$ could in principle limit the nuclear spin coherence time $T_2$, it is typically orders of magnitude larger than the $T_2$ measured here \cite{castner_raman_1963}, which is confirmed by the polarization dynamics discussed above. However, we suspect that electron spin flip-flop processes limit the nuclear spin coherence time in this sample because of the rather large phosphorus concentration.

Coming back to the application of our experiments to $^\mathrm{nat}$Si, we first measure the photoconductive spectrum at $B=300$ mT and $T=6$ K. As shown in Fig.~\ref{fig:natSi}, the six DBE peaks are observed but overlap significantly. Still, we are able to measure a pulsed AEDMR spectrum (Fig.~\ref{fig:natSi}(b)), as well as Rabi oscillation of the electron spin (Fig.~\ref{fig:natSi}(c)). The difference of the amplitudes of the hyperfine-split resonances is again due to the slow Overhauser polarization of the nuclear spins \cite{dluhy_switchable_2015}. Since the range of the piezo-tuning of our laser is not sufficient to achieve an effective pulsing of the laser, a mechanical shutter should be used to realize more complex experiments, such as ENDOR. Still, these proof-of-principle measurements show that optically assisted ENDOR experiments are also feasible in $^\mathrm{nat}$Si.

In summary, we have presented Auger-electron detected measurements of the electron and nuclear spin of phosphorus donors in $^{28}$Si. Because of the long electron spin dephasing times in this material, AEDMR experiments can be realized at very low mw powers, which allowed us to realize a broadband detection using the antenna of an off-resonant dielectric resonator.
We have further demonstrated the detection and manipulation of the nuclear spin state using Auger-electron-detected ENDOR, which does not rely on the optical selectivity on $m_I$. This approach significantly lowers the requirements on the linewidth of the laser used for excitation, on strain and isotopic purity of the sample, and on the experimental temperature compared to Auger-electron-detected NMR experiments.
Also, we have shown the feasibility of AEDMR of phosphorus in silicon with natural isotope composition, which means that the discussed ENDOR technique can enable the control of the $^{31}$P nuclear spin in such samples as well. The resulting very large polarization can also be transferred to $^{29}$Si nuclear spins to enhance the NMR signal in Si nanoparticle MRI agents, either by ENDOR followed by inter-nuclei relaxation \cite{dluhy_switchable_2015}, or by a direct transfer of the electron spin polarization using dynamic nuclear polarization \cite{hartmann_nuclear_1962,abragam_principles_1978,can_time_2015}.

%\todos
\begin{acknowledgments}
This work was financially supported by DFG through SPP 1601 (grant Br 1858/8-2) and the JST-DFG Strategic Cooperative Program on Nanoelectronics, the work at Keio was supported by KAKENHI (S) No. 26220602, JSPS Core-to-Core Program, and the Spintronics Research Network in Japan.
\end{acknowledgments}

\bibliography{bib4}

%merlin.mbs apsrev4-1.bst 2010-07-25 4.21a (PWD, AO, DPC) hacked
%Control: key (0)
%Control: author (8) initials jnrlst
%Control: editor formatted (1) identically to author
%Control: production of article title (-1) disabled
%Control: page (0) single
%Control: year (1) truncated
%Control: production of eprint (0) enabled
\begin{thebibliography}{49}%
\makeatletter
\providecommand \@ifxundefined [1]{%
 \@ifx{#1\undefined}
}%
\providecommand \@ifnum [1]{%
 \ifnum #1\expandafter \@firstoftwo
 \else \expandafter \@secondoftwo
 \fi
}%
\providecommand \@ifx [1]{%
 \ifx #1\expandafter \@firstoftwo
 \else \expandafter \@secondoftwo
 \fi
}%
\providecommand \natexlab [1]{#1}%
\providecommand \enquote  [1]{``#1''}%
\providecommand \bibnamefont  [1]{#1}%
\providecommand \bibfnamefont [1]{#1}%
\providecommand \citenamefont [1]{#1}%
\providecommand \href@noop [0]{\@secondoftwo}%
\providecommand \href [0]{\begingroup \@sanitize@url \@href}%
\providecommand \@href[1]{\@@startlink{#1}\@@href}%
\providecommand \@@href[1]{\endgroup#1\@@endlink}%
\providecommand \@sanitize@url [0]{\catcode `\\12\catcode `\$12\catcode
  `\&12\catcode `\#12\catcode `\^12\catcode `\_12\catcode `\%12\relax}%
\providecommand \@@startlink[1]{}%
\providecommand \@@endlink[0]{}%
\providecommand \url  [0]{\begingroup\@sanitize@url \@url }%
\providecommand \@url [1]{\endgroup\@href {#1}{\urlprefix }}%
\providecommand \urlprefix  [0]{URL }%
\providecommand \Eprint [0]{\href }%
\providecommand \doibase [0]{http://dx.doi.org/}%
\providecommand \selectlanguage [0]{\@gobble}%
\providecommand \bibinfo  [0]{\@secondoftwo}%
\providecommand \bibfield  [0]{\@secondoftwo}%
\providecommand \translation [1]{[#1]}%
\providecommand \BibitemOpen [0]{}%
\providecommand \bibitemStop [0]{}%
\providecommand \bibitemNoStop [0]{.\EOS\space}%
\providecommand \EOS [0]{\spacefactor3000\relax}%
\providecommand \BibitemShut  [1]{\csname bibitem#1\endcsname}%
\let\auto@bib@innerbib\@empty
%</preamble>
\bibitem [{\citenamefont {Stegner}\ \emph {et~al.}(2006)\citenamefont
  {Stegner}, \citenamefont {Boehme}, \citenamefont {Huebl}, \citenamefont
  {Stutzmann}, \citenamefont {Lips},\ and\ \citenamefont
  {Brandt}}]{stegner_electrical_2006}%
  \BibitemOpen
  \bibfield  {author} {\bibinfo {author} {\bibfnamefont {A.~R.}\ \bibnamefont
  {Stegner}}, \bibinfo {author} {\bibfnamefont {C.}~\bibnamefont {Boehme}},
  \bibinfo {author} {\bibfnamefont {H.}~\bibnamefont {Huebl}}, \bibinfo
  {author} {\bibfnamefont {M.}~\bibnamefont {Stutzmann}}, \bibinfo {author}
  {\bibfnamefont {K.}~\bibnamefont {Lips}}, \ and\ \bibinfo {author}
  {\bibfnamefont {M.~S.}\ \bibnamefont {Brandt}},\ }\href {\doibase
  10.1038/nphys465} {\bibfield  {journal} {\bibinfo  {journal} {Nat. Phys.}\
  }\textbf {\bibinfo {volume} {2}},\ \bibinfo {pages} {835} (\bibinfo {year}
  {2006})}\BibitemShut {NoStop}%
\bibitem [{\citenamefont {McCamey}\ \emph {et~al.}(2006)\citenamefont
  {McCamey}, \citenamefont {Huebl}, \citenamefont {Brandt}, \citenamefont
  {Hutchison}, \citenamefont {McCallum}, \citenamefont {Clark},\ and\
  \citenamefont {Hamilton}}]{mccamey_electrically_2006}%
  \BibitemOpen
  \bibfield  {author} {\bibinfo {author} {\bibfnamefont {D.~R.}\ \bibnamefont
  {McCamey}}, \bibinfo {author} {\bibfnamefont {H.}~\bibnamefont {Huebl}},
  \bibinfo {author} {\bibfnamefont {M.~S.}\ \bibnamefont {Brandt}}, \bibinfo
  {author} {\bibfnamefont {W.~D.}\ \bibnamefont {Hutchison}}, \bibinfo {author}
  {\bibfnamefont {J.~C.}\ \bibnamefont {McCallum}}, \bibinfo {author}
  {\bibfnamefont {R.~G.}\ \bibnamefont {Clark}}, \ and\ \bibinfo {author}
  {\bibfnamefont {A.~R.}\ \bibnamefont {Hamilton}},\ }\href {\doibase
  10.1063/1.2358928} {\bibfield  {journal} {\bibinfo  {journal} {J. Appl.
  Phys.}\ }\textbf {\bibinfo {volume} {89}},\ \bibinfo {pages} {182115}
  (\bibinfo {year} {2006})}\BibitemShut {NoStop}%
\bibitem [{\citenamefont {McCamey}\ \emph {et~al.}(2010)\citenamefont
  {McCamey}, \citenamefont {Tol}, \citenamefont {Morley},\ and\ \citenamefont
  {Boehme}}]{mccamey_electronic_2010}%
  \BibitemOpen
  \bibfield  {author} {\bibinfo {author} {\bibfnamefont {D.~R.}\ \bibnamefont
  {McCamey}}, \bibinfo {author} {\bibfnamefont {J.~V.}\ \bibnamefont {Tol}},
  \bibinfo {author} {\bibfnamefont {G.~W.}\ \bibnamefont {Morley}}, \ and\
  \bibinfo {author} {\bibfnamefont {C.}~\bibnamefont {Boehme}},\ }\href
  {\doibase 10.1126/science.1197931} {\bibfield  {journal} {\bibinfo  {journal}
  {Science}\ }\textbf {\bibinfo {volume} {330}},\ \bibinfo {pages} {1652}
  (\bibinfo {year} {2010})}\BibitemShut {NoStop}%
\bibitem [{\citenamefont {Steger}\ \emph {et~al.}(2012)\citenamefont {Steger},
  \citenamefont {Saeedi}, \citenamefont {Thewalt}, \citenamefont {Morton},
  \citenamefont {Riemann}, \citenamefont {Abrosimov}, \citenamefont {Becker},\
  and\ \citenamefont {Pohl}}]{steger_quantum_2012}%
  \BibitemOpen
  \bibfield  {author} {\bibinfo {author} {\bibfnamefont {M.}~\bibnamefont
  {Steger}}, \bibinfo {author} {\bibfnamefont {K.}~\bibnamefont {Saeedi}},
  \bibinfo {author} {\bibfnamefont {M.~L.~W.}\ \bibnamefont {Thewalt}},
  \bibinfo {author} {\bibfnamefont {J.~J.~L.}\ \bibnamefont {Morton}}, \bibinfo
  {author} {\bibfnamefont {H.}~\bibnamefont {Riemann}}, \bibinfo {author}
  {\bibfnamefont {N.~V.}\ \bibnamefont {Abrosimov}}, \bibinfo {author}
  {\bibfnamefont {P.}~\bibnamefont {Becker}}, \ and\ \bibinfo {author}
  {\bibfnamefont {H.-J.}\ \bibnamefont {Pohl}},\ }\href {\doibase
  10.1126/science.1217635} {\bibfield  {journal} {\bibinfo  {journal}
  {Science}\ }\textbf {\bibinfo {volume} {336}},\ \bibinfo {pages} {1280}
  (\bibinfo {year} {2012})}\BibitemShut {NoStop}%
\bibitem [{\citenamefont {Morello}\ \emph {et~al.}(2010)\citenamefont
  {Morello}, \citenamefont {Pla}, \citenamefont {Zwanenburg}, \citenamefont
  {Chan}, \citenamefont {Tan}, \citenamefont {Huebl}, \citenamefont
  {Möttönen}, \citenamefont {Nugroho}, \citenamefont {Yang}, \citenamefont
  {Donkelaar}, \citenamefont {Alves}, \citenamefont {Jamieson}, \citenamefont
  {Escott}, \citenamefont {Hollenberg}, \citenamefont {Clark},\ and\
  \citenamefont {Dzurak}}]{morello_single-shot_2010}%
  \BibitemOpen
  \bibfield  {author} {\bibinfo {author} {\bibfnamefont {A.}~\bibnamefont
  {Morello}}, \bibinfo {author} {\bibfnamefont {J.~J.}\ \bibnamefont {Pla}},
  \bibinfo {author} {\bibfnamefont {F.~A.}\ \bibnamefont {Zwanenburg}},
  \bibinfo {author} {\bibfnamefont {K.~W.}\ \bibnamefont {Chan}}, \bibinfo
  {author} {\bibfnamefont {K.~Y.}\ \bibnamefont {Tan}}, \bibinfo {author}
  {\bibfnamefont {H.}~\bibnamefont {Huebl}}, \bibinfo {author} {\bibfnamefont
  {M.}~\bibnamefont {Möttönen}}, \bibinfo {author} {\bibfnamefont {C.~D.}\
  \bibnamefont {Nugroho}}, \bibinfo {author} {\bibfnamefont {C.}~\bibnamefont
  {Yang}}, \bibinfo {author} {\bibfnamefont {J.~A.~v.}\ \bibnamefont
  {Donkelaar}}, \bibinfo {author} {\bibfnamefont {A.~D.~C.}\ \bibnamefont
  {Alves}}, \bibinfo {author} {\bibfnamefont {D.~N.}\ \bibnamefont {Jamieson}},
  \bibinfo {author} {\bibfnamefont {C.~C.}\ \bibnamefont {Escott}}, \bibinfo
  {author} {\bibfnamefont {L.~C.~L.}\ \bibnamefont {Hollenberg}}, \bibinfo
  {author} {\bibfnamefont {R.~G.}\ \bibnamefont {Clark}}, \ and\ \bibinfo
  {author} {\bibfnamefont {A.~S.}\ \bibnamefont {Dzurak}},\ }\href {\doibase
  10.1038/nature09392} {\bibfield  {journal} {\bibinfo  {journal} {Nature}\
  }\textbf {\bibinfo {volume} {467}},\ \bibinfo {pages} {687} (\bibinfo {year}
  {2010})}\BibitemShut {NoStop}%
\bibitem [{\citenamefont {Pla}\ \emph {et~al.}(2013)\citenamefont {Pla},
  \citenamefont {Tan}, \citenamefont {Dehollain}, \citenamefont {Lim},
  \citenamefont {Morton}, \citenamefont {Zwanenburg}, \citenamefont {Jamieson},
  \citenamefont {Dzurak},\ and\ \citenamefont
  {Morello}}]{pla_high-fidelity_2013}%
  \BibitemOpen
  \bibfield  {author} {\bibinfo {author} {\bibfnamefont {J.~J.}\ \bibnamefont
  {Pla}}, \bibinfo {author} {\bibfnamefont {K.~Y.}\ \bibnamefont {Tan}},
  \bibinfo {author} {\bibfnamefont {J.~P.}\ \bibnamefont {Dehollain}}, \bibinfo
  {author} {\bibfnamefont {W.~H.}\ \bibnamefont {Lim}}, \bibinfo {author}
  {\bibfnamefont {J.~J.~L.}\ \bibnamefont {Morton}}, \bibinfo {author}
  {\bibfnamefont {F.~A.}\ \bibnamefont {Zwanenburg}}, \bibinfo {author}
  {\bibfnamefont {D.~N.}\ \bibnamefont {Jamieson}}, \bibinfo {author}
  {\bibfnamefont {A.~S.}\ \bibnamefont {Dzurak}}, \ and\ \bibinfo {author}
  {\bibfnamefont {A.}~\bibnamefont {Morello}},\ }\href {\doibase
  10.1038/nature12011} {\bibfield  {journal} {\bibinfo  {journal} {Nature}\
  }\textbf {\bibinfo {volume} {496}},\ \bibinfo {pages} {334} (\bibinfo {year}
  {2013})}\BibitemShut {NoStop}%
\bibitem [{\citenamefont {Morishita}\ \emph {et~al.}(2009)\citenamefont
  {Morishita}, \citenamefont {Vlasenko}, \citenamefont {Tanaka}, \citenamefont
  {Semba}, \citenamefont {Sawano}, \citenamefont {Shiraki}, \citenamefont
  {Eto},\ and\ \citenamefont {Itoh}}]{morishita_electrical_2009}%
  \BibitemOpen
  \bibfield  {author} {\bibinfo {author} {\bibfnamefont {H.}~\bibnamefont
  {Morishita}}, \bibinfo {author} {\bibfnamefont {L.~S.}\ \bibnamefont
  {Vlasenko}}, \bibinfo {author} {\bibfnamefont {H.}~\bibnamefont {Tanaka}},
  \bibinfo {author} {\bibfnamefont {K.}~\bibnamefont {Semba}}, \bibinfo
  {author} {\bibfnamefont {K.}~\bibnamefont {Sawano}}, \bibinfo {author}
  {\bibfnamefont {Y.}~\bibnamefont {Shiraki}}, \bibinfo {author} {\bibfnamefont
  {M.}~\bibnamefont {Eto}}, \ and\ \bibinfo {author} {\bibfnamefont {K.~M.}\
  \bibnamefont {Itoh}},\ }\href {\doibase 10.1103/PhysRevB.80.205206}
  {\bibfield  {journal} {\bibinfo  {journal} {Phys. Rev. B}\ }\textbf {\bibinfo
  {volume} {80}},\ \bibinfo {pages} {205206} (\bibinfo {year}
  {2009})}\BibitemShut {NoStop}%
\bibitem [{\citenamefont {Franke}\ \emph {et~al.}(2014)\citenamefont {Franke},
  \citenamefont {Hoehne}, \citenamefont {Vlasenko}, \citenamefont {Itoh},\ and\
  \citenamefont {Brandt}}]{franke_spin-dependent_2014-1}%
  \BibitemOpen
  \bibfield  {author} {\bibinfo {author} {\bibfnamefont {D.~P.}\ \bibnamefont
  {Franke}}, \bibinfo {author} {\bibfnamefont {F.}~\bibnamefont {Hoehne}},
  \bibinfo {author} {\bibfnamefont {L.~S.}\ \bibnamefont {Vlasenko}}, \bibinfo
  {author} {\bibfnamefont {K.~M.}\ \bibnamefont {Itoh}}, \ and\ \bibinfo
  {author} {\bibfnamefont {M.~S.}\ \bibnamefont {Brandt}},\ }\href {\doibase
  10.1103/PhysRevB.89.195207} {\bibfield  {journal} {\bibinfo  {journal} {Phys.
  Rev. B}\ }\textbf {\bibinfo {volume} {89}},\ \bibinfo {pages} {195207}
  (\bibinfo {year} {2014})}\BibitemShut {NoStop}%
\bibitem [{\citenamefont {Dreher}\ \emph {et~al.}(2015)\citenamefont {Dreher},
  \citenamefont {Hoehne}, \citenamefont {Morishita}, \citenamefont {Huebl},
  \citenamefont {Stutzmann}, \citenamefont {Itoh},\ and\ \citenamefont
  {Brandt}}]{dreher_pulsed_2015}%
  \BibitemOpen
  \bibfield  {author} {\bibinfo {author} {\bibfnamefont {L.}~\bibnamefont
  {Dreher}}, \bibinfo {author} {\bibfnamefont {F.}~\bibnamefont {Hoehne}},
  \bibinfo {author} {\bibfnamefont {H.}~\bibnamefont {Morishita}}, \bibinfo
  {author} {\bibfnamefont {H.}~\bibnamefont {Huebl}}, \bibinfo {author}
  {\bibfnamefont {M.}~\bibnamefont {Stutzmann}}, \bibinfo {author}
  {\bibfnamefont {K.~M.}\ \bibnamefont {Itoh}}, \ and\ \bibinfo {author}
  {\bibfnamefont {M.~S.}\ \bibnamefont {Brandt}},\ }\href {\doibase
  10.1103/PhysRevB.91.075314} {\bibfield  {journal} {\bibinfo  {journal} {Phys.
  Rev. B}\ }\textbf {\bibinfo {volume} {91}},\ \bibinfo {pages} {075314}
  (\bibinfo {year} {2015})}\BibitemShut {NoStop}%
\bibitem [{\citenamefont {Tyryshkin}\ \emph {et~al.}(2003)\citenamefont
  {Tyryshkin}, \citenamefont {Lyon}, \citenamefont {Astashkin},\ and\
  \citenamefont {Raitsimring}}]{tyryshkin_electron_2003}%
  \BibitemOpen
  \bibfield  {author} {\bibinfo {author} {\bibfnamefont {A.~M.}\ \bibnamefont
  {Tyryshkin}}, \bibinfo {author} {\bibfnamefont {S.~A.}\ \bibnamefont {Lyon}},
  \bibinfo {author} {\bibfnamefont {A.~V.}\ \bibnamefont {Astashkin}}, \ and\
  \bibinfo {author} {\bibfnamefont {A.~M.}\ \bibnamefont {Raitsimring}},\
  }\href {\doibase 10.1103/PhysRevB.68.193207} {\bibfield  {journal} {\bibinfo
  {journal} {Phys. Rev. B}\ }\textbf {\bibinfo {volume} {68}},\ \bibinfo
  {pages} {193207} (\bibinfo {year} {2003})}\BibitemShut {NoStop}%
\bibitem [{\citenamefont {Tyryshkin}\ \emph {et~al.}(2012)\citenamefont
  {Tyryshkin}, \citenamefont {Tojo}, \citenamefont {Morton}, \citenamefont
  {Riemann}, \citenamefont {Abrosimov}, \citenamefont {Becker}, \citenamefont
  {Pohl}, \citenamefont {Schenkel}, \citenamefont {Thewalt}, \citenamefont
  {Itoh},\ and\ \citenamefont {Lyon}}]{tyryshkin_electron_2012}%
  \BibitemOpen
  \bibfield  {author} {\bibinfo {author} {\bibfnamefont {A.~M.}\ \bibnamefont
  {Tyryshkin}}, \bibinfo {author} {\bibfnamefont {S.}~\bibnamefont {Tojo}},
  \bibinfo {author} {\bibfnamefont {J.~J.~L.}\ \bibnamefont {Morton}}, \bibinfo
  {author} {\bibfnamefont {H.}~\bibnamefont {Riemann}}, \bibinfo {author}
  {\bibfnamefont {N.~V.}\ \bibnamefont {Abrosimov}}, \bibinfo {author}
  {\bibfnamefont {P.}~\bibnamefont {Becker}}, \bibinfo {author} {\bibfnamefont
  {H.-J.}\ \bibnamefont {Pohl}}, \bibinfo {author} {\bibfnamefont
  {T.}~\bibnamefont {Schenkel}}, \bibinfo {author} {\bibfnamefont {M.~L.~W.}\
  \bibnamefont {Thewalt}}, \bibinfo {author} {\bibfnamefont {K.~M.}\
  \bibnamefont {Itoh}}, \ and\ \bibinfo {author} {\bibfnamefont {S.~A.}\
  \bibnamefont {Lyon}},\ }\href {\doibase 10.1038/nmat3182} {\bibfield
  {journal} {\bibinfo  {journal} {Nat. Mater.}\ }\textbf {\bibinfo {volume}
  {11}},\ \bibinfo {pages} {143} (\bibinfo {year} {2012})}\BibitemShut
  {NoStop}%
\bibitem [{\citenamefont {Saeedi}\ \emph {et~al.}(2013)\citenamefont {Saeedi},
  \citenamefont {Simmons}, \citenamefont {Salvail}, \citenamefont {Dluhy},
  \citenamefont {Riemann}, \citenamefont {Abrosimov}, \citenamefont {Becker},
  \citenamefont {Pohl}, \citenamefont {Morton},\ and\ \citenamefont
  {Thewalt}}]{saeedi_room-temperature_2013}%
  \BibitemOpen
  \bibfield  {author} {\bibinfo {author} {\bibfnamefont {K.}~\bibnamefont
  {Saeedi}}, \bibinfo {author} {\bibfnamefont {S.}~\bibnamefont {Simmons}},
  \bibinfo {author} {\bibfnamefont {J.~Z.}\ \bibnamefont {Salvail}}, \bibinfo
  {author} {\bibfnamefont {P.}~\bibnamefont {Dluhy}}, \bibinfo {author}
  {\bibfnamefont {H.}~\bibnamefont {Riemann}}, \bibinfo {author} {\bibfnamefont
  {N.~V.}\ \bibnamefont {Abrosimov}}, \bibinfo {author} {\bibfnamefont
  {P.}~\bibnamefont {Becker}}, \bibinfo {author} {\bibfnamefont {H.-J.}\
  \bibnamefont {Pohl}}, \bibinfo {author} {\bibfnamefont {J.~J.~L.}\
  \bibnamefont {Morton}}, \ and\ \bibinfo {author} {\bibfnamefont {M.~L.~W.}\
  \bibnamefont {Thewalt}},\ }\href {\doibase 10.1126/science.1239584}
  {\bibfield  {journal} {\bibinfo  {journal} {Science}\ }\textbf {\bibinfo
  {volume} {342}},\ \bibinfo {pages} {830} (\bibinfo {year}
  {2013})}\BibitemShut {NoStop}%
\bibitem [{\citenamefont {Itoh}\ and\ \citenamefont
  {Watanabe}(2014)}]{itoh_isotope_2014}%
  \BibitemOpen
  \bibfield  {author} {\bibinfo {author} {\bibfnamefont {K.~M.}\ \bibnamefont
  {Itoh}}\ and\ \bibinfo {author} {\bibfnamefont {H.}~\bibnamefont
  {Watanabe}},\ }\href {\doibase 10.1557/mrc.2014.32} {\bibfield  {journal}
  {\bibinfo  {journal} {MRS Commun.}\ }\textbf {\bibinfo {volume} {4}},\
  \bibinfo {pages} {143} (\bibinfo {year} {2014})}\BibitemShut {NoStop}%
\bibitem [{\citenamefont {Bradbury}\ \emph {et~al.}(2006)\citenamefont
  {Bradbury}, \citenamefont {Tyryshkin}, \citenamefont {Sabouret},
  \citenamefont {Bokor}, \citenamefont {Schenkel},\ and\ \citenamefont
  {Lyon}}]{bradbury_stark_2006}%
  \BibitemOpen
  \bibfield  {author} {\bibinfo {author} {\bibfnamefont {F.~R.}\ \bibnamefont
  {Bradbury}}, \bibinfo {author} {\bibfnamefont {A.~M.}\ \bibnamefont
  {Tyryshkin}}, \bibinfo {author} {\bibfnamefont {G.}~\bibnamefont {Sabouret}},
  \bibinfo {author} {\bibfnamefont {J.}~\bibnamefont {Bokor}}, \bibinfo
  {author} {\bibfnamefont {T.}~\bibnamefont {Schenkel}}, \ and\ \bibinfo
  {author} {\bibfnamefont {S.~A.}\ \bibnamefont {Lyon}},\ }\href {\doibase
  10.1103/PhysRevLett.97.176404} {\bibfield  {journal} {\bibinfo  {journal}
  {Phys. Rev. Lett.}\ }\textbf {\bibinfo {volume} {97}},\ \bibinfo {pages}
  {176404} (\bibinfo {year} {2006})}\BibitemShut {NoStop}%
\bibitem [{\citenamefont {Lo}\ \emph {et~al.}(2014)\citenamefont {Lo},
  \citenamefont {Simmons}, \citenamefont {Nardo}, \citenamefont {Weis},
  \citenamefont {Tyryshkin}, \citenamefont {Meijer}, \citenamefont {Rogalla},
  \citenamefont {Lyon}, \citenamefont {Bokor}, \citenamefont {Schenkel},\ and\
  \citenamefont {Morton}}]{lo_stark_2014}%
  \BibitemOpen
  \bibfield  {author} {\bibinfo {author} {\bibfnamefont {C.~C.}\ \bibnamefont
  {Lo}}, \bibinfo {author} {\bibfnamefont {S.}~\bibnamefont {Simmons}},
  \bibinfo {author} {\bibfnamefont {R.~L.}\ \bibnamefont {Nardo}}, \bibinfo
  {author} {\bibfnamefont {C.~D.}\ \bibnamefont {Weis}}, \bibinfo {author}
  {\bibfnamefont {A.~M.}\ \bibnamefont {Tyryshkin}}, \bibinfo {author}
  {\bibfnamefont {J.}~\bibnamefont {Meijer}}, \bibinfo {author} {\bibfnamefont
  {D.}~\bibnamefont {Rogalla}}, \bibinfo {author} {\bibfnamefont {S.~A.}\
  \bibnamefont {Lyon}}, \bibinfo {author} {\bibfnamefont {J.}~\bibnamefont
  {Bokor}}, \bibinfo {author} {\bibfnamefont {T.}~\bibnamefont {Schenkel}}, \
  and\ \bibinfo {author} {\bibfnamefont {J.~J.~L.}\ \bibnamefont {Morton}},\
  }\href {\doibase 10.1063/1.4876175} {\bibfield  {journal} {\bibinfo
  {journal} {Appl. Phys. Lett.}\ }\textbf {\bibinfo {volume} {104}},\ \bibinfo
  {pages} {193502} (\bibinfo {year} {2014})}\BibitemShut {NoStop}%
\bibitem [{\citenamefont {Wolfowicz}\ \emph {et~al.}(2014)\citenamefont
  {Wolfowicz}, \citenamefont {Urdampilleta}, \citenamefont {Thewalt},
  \citenamefont {Riemann}, \citenamefont {Abrosimov}, \citenamefont {Becker},
  \citenamefont {Pohl},\ and\ \citenamefont
  {Morton}}]{wolfowicz_conditional_2014}%
  \BibitemOpen
  \bibfield  {author} {\bibinfo {author} {\bibfnamefont {G.}~\bibnamefont
  {Wolfowicz}}, \bibinfo {author} {\bibfnamefont {M.}~\bibnamefont
  {Urdampilleta}}, \bibinfo {author} {\bibfnamefont {M.~L.~W.}\ \bibnamefont
  {Thewalt}}, \bibinfo {author} {\bibfnamefont {H.}~\bibnamefont {Riemann}},
  \bibinfo {author} {\bibfnamefont {N.~V.}\ \bibnamefont {Abrosimov}}, \bibinfo
  {author} {\bibfnamefont {P.}~\bibnamefont {Becker}}, \bibinfo {author}
  {\bibfnamefont {H.-J.}\ \bibnamefont {Pohl}}, \ and\ \bibinfo {author}
  {\bibfnamefont {J.~J.~L.}\ \bibnamefont {Morton}},\ }\href {\doibase
  10.1103/PhysRevLett.113.157601} {\bibfield  {journal} {\bibinfo  {journal}
  {Phys. Rev. Lett.}\ }\textbf {\bibinfo {volume} {113}},\ \bibinfo {pages}
  {157601} (\bibinfo {year} {2014})}\BibitemShut {NoStop}%
\bibitem [{\citenamefont {Laucht}\ \emph {et~al.}(2015)\citenamefont {Laucht},
  \citenamefont {Muhonen}, \citenamefont {Mohiyaddin}, \citenamefont {Kalra},
  \citenamefont {Dehollain}, \citenamefont {Freer}, \citenamefont {Hudson},
  \citenamefont {Veldhorst}, \citenamefont {Rahman}, \citenamefont {Klimeck},
  \citenamefont {Itoh}, \citenamefont {Jamieson}, \citenamefont {McCallum},
  \citenamefont {Dzurak},\ and\ \citenamefont
  {Morello}}]{laucht_electrically_2015}%
  \BibitemOpen
  \bibfield  {author} {\bibinfo {author} {\bibfnamefont {A.}~\bibnamefont
  {Laucht}}, \bibinfo {author} {\bibfnamefont {J.~T.}\ \bibnamefont {Muhonen}},
  \bibinfo {author} {\bibfnamefont {F.~A.}\ \bibnamefont {Mohiyaddin}},
  \bibinfo {author} {\bibfnamefont {R.}~\bibnamefont {Kalra}}, \bibinfo
  {author} {\bibfnamefont {J.~P.}\ \bibnamefont {Dehollain}}, \bibinfo {author}
  {\bibfnamefont {S.}~\bibnamefont {Freer}}, \bibinfo {author} {\bibfnamefont
  {F.~E.}\ \bibnamefont {Hudson}}, \bibinfo {author} {\bibfnamefont
  {M.}~\bibnamefont {Veldhorst}}, \bibinfo {author} {\bibfnamefont
  {R.}~\bibnamefont {Rahman}}, \bibinfo {author} {\bibfnamefont
  {G.}~\bibnamefont {Klimeck}}, \bibinfo {author} {\bibfnamefont {K.~M.}\
  \bibnamefont {Itoh}}, \bibinfo {author} {\bibfnamefont {D.~N.}\ \bibnamefont
  {Jamieson}}, \bibinfo {author} {\bibfnamefont {J.~C.}\ \bibnamefont
  {McCallum}}, \bibinfo {author} {\bibfnamefont {A.~S.}\ \bibnamefont
  {Dzurak}}, \ and\ \bibinfo {author} {\bibfnamefont {A.}~\bibnamefont
  {Morello}},\ }\href {\doibase 10.1126/sciadv.1500022} {\bibfield  {journal}
  {\bibinfo  {journal} {Sci. Adv.}\ }\textbf {\bibinfo {volume} {1}},\ \bibinfo
  {pages} {e1500022} (\bibinfo {year} {2015})}\BibitemShut {NoStop}%
\bibitem [{\citenamefont {Hoehne}\ \emph {et~al.}(2013)\citenamefont {Hoehne},
  \citenamefont {Dreher}, \citenamefont {Suckert}, \citenamefont {Franke},
  \citenamefont {Stutzmann},\ and\ \citenamefont {Brandt}}]{hoehne_time_2013}%
  \BibitemOpen
  \bibfield  {author} {\bibinfo {author} {\bibfnamefont {F.}~\bibnamefont
  {Hoehne}}, \bibinfo {author} {\bibfnamefont {L.}~\bibnamefont {Dreher}},
  \bibinfo {author} {\bibfnamefont {M.}~\bibnamefont {Suckert}}, \bibinfo
  {author} {\bibfnamefont {D.~P.}\ \bibnamefont {Franke}}, \bibinfo {author}
  {\bibfnamefont {M.}~\bibnamefont {Stutzmann}}, \ and\ \bibinfo {author}
  {\bibfnamefont {M.~S.}\ \bibnamefont {Brandt}},\ }\href {\doibase
  10.1103/PhysRevB.88.155301} {\bibfield  {journal} {\bibinfo  {journal} {Phys.
  Rev. B}\ }\textbf {\bibinfo {volume} {88}},\ \bibinfo {pages} {155301}
  (\bibinfo {year} {2013})}\BibitemShut {NoStop}%
\bibitem [{\citenamefont {Dreher}\ \emph {et~al.}(2012)\citenamefont {Dreher},
  \citenamefont {Hoehne}, \citenamefont {Stutzmann},\ and\ \citenamefont
  {Brandt}}]{dreher_nuclear_2012}%
  \BibitemOpen
  \bibfield  {author} {\bibinfo {author} {\bibfnamefont {L.}~\bibnamefont
  {Dreher}}, \bibinfo {author} {\bibfnamefont {F.}~\bibnamefont {Hoehne}},
  \bibinfo {author} {\bibfnamefont {M.}~\bibnamefont {Stutzmann}}, \ and\
  \bibinfo {author} {\bibfnamefont {M.~S.}\ \bibnamefont {Brandt}},\ }\href
  {\doibase 10.1103/PhysRevLett.108.027602} {\bibfield  {journal} {\bibinfo
  {journal} {Phys. Rev. Lett.}\ }\textbf {\bibinfo {volume} {108}},\ \bibinfo
  {pages} {027602} (\bibinfo {year} {2012})}\BibitemShut {NoStop}%
\bibitem [{\citenamefont {Hoehne}\ \emph {et~al.}(2015)\citenamefont {Hoehne},
  \citenamefont {Dreher}, \citenamefont {Franke}, \citenamefont {Stutzmann},
  \citenamefont {Vlasenko}, \citenamefont {Itoh},\ and\ \citenamefont
  {Brandt}}]{hoehne_submillisecond_2015}%
  \BibitemOpen
  \bibfield  {author} {\bibinfo {author} {\bibfnamefont {F.}~\bibnamefont
  {Hoehne}}, \bibinfo {author} {\bibfnamefont {L.}~\bibnamefont {Dreher}},
  \bibinfo {author} {\bibfnamefont {D.~P.}\ \bibnamefont {Franke}}, \bibinfo
  {author} {\bibfnamefont {M.}~\bibnamefont {Stutzmann}}, \bibinfo {author}
  {\bibfnamefont {L.~S.}\ \bibnamefont {Vlasenko}}, \bibinfo {author}
  {\bibfnamefont {K.~M.}\ \bibnamefont {Itoh}}, \ and\ \bibinfo {author}
  {\bibfnamefont {M.~S.}\ \bibnamefont {Brandt}},\ }\href {\doibase
  10.1103/PhysRevLett.114.117602} {\bibfield  {journal} {\bibinfo  {journal}
  {Phys. Rev. Lett.}\ }\textbf {\bibinfo {volume} {114}},\ \bibinfo {pages}
  {117602} (\bibinfo {year} {2015})}\BibitemShut {NoStop}%
\bibitem [{\citenamefont {Karaiskaj}\ \emph {et~al.}(2001)\citenamefont
  {Karaiskaj}, \citenamefont {Thewalt}, \citenamefont {Ruf}, \citenamefont
  {Cardona}, \citenamefont {Pohl}, \citenamefont {Deviatych}, \citenamefont
  {Sennikov},\ and\ \citenamefont
  {Riemann}}]{karaiskaj_photoluminescence_2001}%
  \BibitemOpen
  \bibfield  {author} {\bibinfo {author} {\bibfnamefont {D.}~\bibnamefont
  {Karaiskaj}}, \bibinfo {author} {\bibfnamefont {M.~L.~W.}\ \bibnamefont
  {Thewalt}}, \bibinfo {author} {\bibfnamefont {T.}~\bibnamefont {Ruf}},
  \bibinfo {author} {\bibfnamefont {M.}~\bibnamefont {Cardona}}, \bibinfo
  {author} {\bibfnamefont {H.-J.}\ \bibnamefont {Pohl}}, \bibinfo {author}
  {\bibfnamefont {G.~G.}\ \bibnamefont {Deviatych}}, \bibinfo {author}
  {\bibfnamefont {P.~G.}\ \bibnamefont {Sennikov}}, \ and\ \bibinfo {author}
  {\bibfnamefont {H.}~\bibnamefont {Riemann}},\ }\href {\doibase
  10.1103/PhysRevLett.86.6010} {\bibfield  {journal} {\bibinfo  {journal}
  {Phys. Rev. Lett.}\ }\textbf {\bibinfo {volume} {86}},\ \bibinfo {pages}
  {6010} (\bibinfo {year} {2001})}\BibitemShut {NoStop}%
\bibitem [{\citenamefont {Yang}\ \emph {et~al.}(2006)\citenamefont {Yang},
  \citenamefont {Steger}, \citenamefont {Karaiskaj}, \citenamefont {Thewalt},
  \citenamefont {Cardona}, \citenamefont {Itoh}, \citenamefont {Riemann},
  \citenamefont {Abrosimov}, \citenamefont {Churbanov}, \citenamefont {Gusev},
  \citenamefont {Bulanov}, \citenamefont {Kaliteevskii}, \citenamefont
  {Godisov}, \citenamefont {Becker}, \citenamefont {Pohl}, \citenamefont
  {Ager},\ and\ \citenamefont {Haller}}]{yang_optical_2006}%
  \BibitemOpen
  \bibfield  {author} {\bibinfo {author} {\bibfnamefont {A.}~\bibnamefont
  {Yang}}, \bibinfo {author} {\bibfnamefont {M.}~\bibnamefont {Steger}},
  \bibinfo {author} {\bibfnamefont {D.}~\bibnamefont {Karaiskaj}}, \bibinfo
  {author} {\bibfnamefont {M.~L.~W.}\ \bibnamefont {Thewalt}}, \bibinfo
  {author} {\bibfnamefont {M.}~\bibnamefont {Cardona}}, \bibinfo {author}
  {\bibfnamefont {K.~M.}\ \bibnamefont {Itoh}}, \bibinfo {author}
  {\bibfnamefont {H.}~\bibnamefont {Riemann}}, \bibinfo {author} {\bibfnamefont
  {N.~V.}\ \bibnamefont {Abrosimov}}, \bibinfo {author} {\bibfnamefont {M.~F.}\
  \bibnamefont {Churbanov}}, \bibinfo {author} {\bibfnamefont {A.~V.}\
  \bibnamefont {Gusev}}, \bibinfo {author} {\bibfnamefont {A.~D.}\ \bibnamefont
  {Bulanov}}, \bibinfo {author} {\bibfnamefont {A.~K.}\ \bibnamefont
  {Kaliteevskii}}, \bibinfo {author} {\bibfnamefont {O.~N.}\ \bibnamefont
  {Godisov}}, \bibinfo {author} {\bibfnamefont {P.}~\bibnamefont {Becker}},
  \bibinfo {author} {\bibfnamefont {H.-J.}\ \bibnamefont {Pohl}}, \bibinfo
  {author} {\bibfnamefont {J.~W.}\ \bibnamefont {Ager}}, \ and\ \bibinfo
  {author} {\bibfnamefont {E.~E.}\ \bibnamefont {Haller}},\ }\href {\doibase
  10.1103/PhysRevLett.97.227401} {\bibfield  {journal} {\bibinfo  {journal}
  {Phys. Rev. Lett.}\ }\textbf {\bibinfo {volume} {97}},\ \bibinfo {pages}
  {227401} (\bibinfo {year} {2006})}\BibitemShut {NoStop}%
\bibitem [{\citenamefont {Yang}\ \emph {et~al.}(2009)\citenamefont {Yang},
  \citenamefont {Steger}, \citenamefont {Sekiguchi}, \citenamefont {Thewalt},
  \citenamefont {Ladd}, \citenamefont {Itoh}, \citenamefont {Riemann},
  \citenamefont {Abrosimov}, \citenamefont {Becker},\ and\ \citenamefont
  {Pohl}}]{yang_simultaneous_2009}%
  \BibitemOpen
  \bibfield  {author} {\bibinfo {author} {\bibfnamefont {A.}~\bibnamefont
  {Yang}}, \bibinfo {author} {\bibfnamefont {M.}~\bibnamefont {Steger}},
  \bibinfo {author} {\bibfnamefont {T.}~\bibnamefont {Sekiguchi}}, \bibinfo
  {author} {\bibfnamefont {M.~L.~W.}\ \bibnamefont {Thewalt}}, \bibinfo
  {author} {\bibfnamefont {T.~D.}\ \bibnamefont {Ladd}}, \bibinfo {author}
  {\bibfnamefont {K.~M.}\ \bibnamefont {Itoh}}, \bibinfo {author}
  {\bibfnamefont {H.}~\bibnamefont {Riemann}}, \bibinfo {author} {\bibfnamefont
  {N.~V.}\ \bibnamefont {Abrosimov}}, \bibinfo {author} {\bibfnamefont
  {P.}~\bibnamefont {Becker}}, \ and\ \bibinfo {author} {\bibfnamefont {H.-J.}\
  \bibnamefont {Pohl}},\ }\href {\doibase 10.1103/PhysRevLett.102.257401}
  {\bibfield  {journal} {\bibinfo  {journal} {Phys. Rev. Lett.}\ }\textbf
  {\bibinfo {volume} {102}},\ \bibinfo {pages} {257401} (\bibinfo {year}
  {2009})}\BibitemShut {NoStop}%
\bibitem [{\citenamefont {Steger}\ \emph {et~al.}(2011)\citenamefont {Steger},
  \citenamefont {Sekiguchi}, \citenamefont {Yang}, \citenamefont {Saeedi},
  \citenamefont {Hayden}, \citenamefont {Thewalt}, \citenamefont {Itoh},
  \citenamefont {Riemann}, \citenamefont {Abrosimov}, \citenamefont {Becker},\
  and\ \citenamefont {Pohl}}]{steger_optically-detected_2011}%
  \BibitemOpen
  \bibfield  {author} {\bibinfo {author} {\bibfnamefont {M.}~\bibnamefont
  {Steger}}, \bibinfo {author} {\bibfnamefont {T.}~\bibnamefont {Sekiguchi}},
  \bibinfo {author} {\bibfnamefont {A.}~\bibnamefont {Yang}}, \bibinfo {author}
  {\bibfnamefont {K.}~\bibnamefont {Saeedi}}, \bibinfo {author} {\bibfnamefont
  {M.~E.}\ \bibnamefont {Hayden}}, \bibinfo {author} {\bibfnamefont {M.~L.~W.}\
  \bibnamefont {Thewalt}}, \bibinfo {author} {\bibfnamefont {K.~M.}\
  \bibnamefont {Itoh}}, \bibinfo {author} {\bibfnamefont {H.}~\bibnamefont
  {Riemann}}, \bibinfo {author} {\bibfnamefont {N.~V.}\ \bibnamefont
  {Abrosimov}}, \bibinfo {author} {\bibfnamefont {P.}~\bibnamefont {Becker}}, \
  and\ \bibinfo {author} {\bibfnamefont {H.-J.}\ \bibnamefont {Pohl}},\ }\href
  {\doibase 10.1063/1.3577614} {\bibfield  {journal} {\bibinfo  {journal} {J.
  Appl. Phys.}\ }\textbf {\bibinfo {volume} {109}},\ \bibinfo {pages} {102411}
  (\bibinfo {year} {2011})}\BibitemShut {NoStop}%
\bibitem [{\citenamefont {Salvail}\ \emph {et~al.}(2015)\citenamefont
  {Salvail}, \citenamefont {Dluhy}, \citenamefont {Morse}, \citenamefont
  {Szech}, \citenamefont {Saeedi}, \citenamefont {Huber}, \citenamefont
  {Riemann}, \citenamefont {Abrosimov}, \citenamefont {Becker}, \citenamefont
  {Pohl},\ and\ \citenamefont {Thewalt}}]{salvail_optically_2015}%
  \BibitemOpen
  \bibfield  {author} {\bibinfo {author} {\bibfnamefont {J.~Z.}\ \bibnamefont
  {Salvail}}, \bibinfo {author} {\bibfnamefont {P.}~\bibnamefont {Dluhy}},
  \bibinfo {author} {\bibfnamefont {K.~J.}\ \bibnamefont {Morse}}, \bibinfo
  {author} {\bibfnamefont {M.}~\bibnamefont {Szech}}, \bibinfo {author}
  {\bibfnamefont {K.}~\bibnamefont {Saeedi}}, \bibinfo {author} {\bibfnamefont
  {J.}~\bibnamefont {Huber}}, \bibinfo {author} {\bibfnamefont
  {H.}~\bibnamefont {Riemann}}, \bibinfo {author} {\bibfnamefont {N.~V.}\
  \bibnamefont {Abrosimov}}, \bibinfo {author} {\bibfnamefont {P.}~\bibnamefont
  {Becker}}, \bibinfo {author} {\bibfnamefont {H.-J.}\ \bibnamefont {Pohl}}, \
  and\ \bibinfo {author} {\bibfnamefont {M.~L.~W.}\ \bibnamefont {Thewalt}},\
  }\href {\doibase 10.1103/PhysRevB.92.195203} {\bibfield  {journal} {\bibinfo
  {journal} {Phys. Rev. B}\ }\textbf {\bibinfo {volume} {92}},\ \bibinfo
  {pages} {195203} (\bibinfo {year} {2015})}\BibitemShut {NoStop}%
\bibitem [{\citenamefont {Kaminskii}\ \emph {et~al.}(1980)\citenamefont
  {Kaminskii}, \citenamefont {Karasyuk},\ and\ \citenamefont
  {Pokrovskii}}]{kaminskii_luminescence_1980}%
  \BibitemOpen
  \bibfield  {author} {\bibinfo {author} {\bibfnamefont {A.~S.}\ \bibnamefont
  {Kaminskii}}, \bibinfo {author} {\bibfnamefont {V.}~\bibnamefont {Karasyuk}},
  \ and\ \bibinfo {author} {\bibfnamefont {Y.~E.}\ \bibnamefont {Pokrovskii}},\
  }\href {http://jetp.ac.ru/cgi-bin/dn/e_052_02_0211.pdf} {\bibfield  {journal}
  {\bibinfo  {journal} {Sov. Phys. JETP,}\ }\textbf {\bibinfo {volume} {52}},\
  \bibinfo {pages} {211} (\bibinfo {year} {1980})}\BibitemShut {NoStop}%
\bibitem [{\citenamefont {Thewalt}\ \emph {et~al.}(2007)\citenamefont
  {Thewalt}, \citenamefont {Yang}, \citenamefont {Steger}, \citenamefont
  {Karaiskaj}, \citenamefont {Cardona}, \citenamefont {Riemann}, \citenamefont
  {Abrosimov}, \citenamefont {Gusev}, \citenamefont {Bulanov}, \citenamefont
  {Kovalev}, \citenamefont {Kaliteevskii}, \citenamefont {Godisov},
  \citenamefont {Becker}, \citenamefont {Pohl}, \citenamefont {Haller},
  \citenamefont {Ager~III},\ and\ \citenamefont {Itoh}}]{thewalt_direct_2007}%
  \BibitemOpen
  \bibfield  {author} {\bibinfo {author} {\bibfnamefont {M.~L.~W.}\
  \bibnamefont {Thewalt}}, \bibinfo {author} {\bibfnamefont {A.}~\bibnamefont
  {Yang}}, \bibinfo {author} {\bibfnamefont {M.}~\bibnamefont {Steger}},
  \bibinfo {author} {\bibfnamefont {D.}~\bibnamefont {Karaiskaj}}, \bibinfo
  {author} {\bibfnamefont {M.}~\bibnamefont {Cardona}}, \bibinfo {author}
  {\bibfnamefont {H.}~\bibnamefont {Riemann}}, \bibinfo {author} {\bibfnamefont
  {N.~V.}\ \bibnamefont {Abrosimov}}, \bibinfo {author} {\bibfnamefont {A.~V.}\
  \bibnamefont {Gusev}}, \bibinfo {author} {\bibfnamefont {A.~D.}\ \bibnamefont
  {Bulanov}}, \bibinfo {author} {\bibfnamefont {I.~D.}\ \bibnamefont
  {Kovalev}}, \bibinfo {author} {\bibfnamefont {A.~K.}\ \bibnamefont
  {Kaliteevskii}}, \bibinfo {author} {\bibfnamefont {O.~N.}\ \bibnamefont
  {Godisov}}, \bibinfo {author} {\bibfnamefont {P.}~\bibnamefont {Becker}},
  \bibinfo {author} {\bibfnamefont {H.~J.}\ \bibnamefont {Pohl}}, \bibinfo
  {author} {\bibfnamefont {E.~E.}\ \bibnamefont {Haller}}, \bibinfo {author}
  {\bibfnamefont {J.~W.}\ \bibnamefont {Ager~III}}, \ and\ \bibinfo {author}
  {\bibfnamefont {K.~M.}\ \bibnamefont {Itoh}},\ }\href {\doibase
  10.1063/1.2723181} {\bibfield  {journal} {\bibinfo  {journal} {J. Appl.
  Phys.}\ }\textbf {\bibinfo {volume} {101}},\ \bibinfo {pages} {081724}
  (\bibinfo {year} {2007})}\BibitemShut {NoStop}%
\bibitem [{\citenamefont {Overhauser}(1953)}]{overhauser_polarization_1953}%
  \BibitemOpen
  \bibfield  {author} {\bibinfo {author} {\bibfnamefont {A.~W.}\ \bibnamefont
  {Overhauser}},\ }\href {\doibase 10.1103/PhysRev.92.411} {\bibfield
  {journal} {\bibinfo  {journal} {Phys. Rev.}\ }\textbf {\bibinfo {volume}
  {92}},\ \bibinfo {pages} {411} (\bibinfo {year} {1953})}\BibitemShut
  {NoStop}%
\bibitem [{\citenamefont {Feher}\ and\ \citenamefont
  {Gere}(1959)}]{feher_electron_1959}%
  \BibitemOpen
  \bibfield  {author} {\bibinfo {author} {\bibfnamefont {G.}~\bibnamefont
  {Feher}}\ and\ \bibinfo {author} {\bibfnamefont {E.~A.}\ \bibnamefont
  {Gere}},\ }\href {\doibase 10.1103/PhysRev.114.1245} {\bibfield  {journal}
  {\bibinfo  {journal} {Phys. Rev.}\ }\textbf {\bibinfo {volume} {114}},\
  \bibinfo {pages} {1245} (\bibinfo {year} {1959})}\BibitemShut {NoStop}%
\bibitem [{\citenamefont {Dluhy}\ \emph {et~al.}(2015)\citenamefont {Dluhy},
  \citenamefont {Salvail}, \citenamefont {Saeedi}, \citenamefont {Thewalt},\
  and\ \citenamefont {Simmons}}]{dluhy_switchable_2015}%
  \BibitemOpen
  \bibfield  {author} {\bibinfo {author} {\bibfnamefont {P.}~\bibnamefont
  {Dluhy}}, \bibinfo {author} {\bibfnamefont {J.~Z.}\ \bibnamefont {Salvail}},
  \bibinfo {author} {\bibfnamefont {K.}~\bibnamefont {Saeedi}}, \bibinfo
  {author} {\bibfnamefont {M.~L.~W.}\ \bibnamefont {Thewalt}}, \ and\ \bibinfo
  {author} {\bibfnamefont {S.}~\bibnamefont {Simmons}},\ }\href {\doibase
  10.1103/PhysRevB.91.195206} {\bibfield  {journal} {\bibinfo  {journal} {Phys.
  Rev. B}\ }\textbf {\bibinfo {volume} {91}},\ \bibinfo {pages} {195206}
  (\bibinfo {year} {2015})}\BibitemShut {NoStop}%
\bibitem [{\citenamefont {Lo}\ \emph {et~al.}(2015)\citenamefont {Lo},
  \citenamefont {Urdampilleta}, \citenamefont {Ross}, \citenamefont
  {Gonzalez-Zalba}, \citenamefont {Mansir}, \citenamefont {Lyon}, \citenamefont
  {Thewalt},\ and\ \citenamefont {Morton}}]{lo_hybrid_2015}%
  \BibitemOpen
  \bibfield  {author} {\bibinfo {author} {\bibfnamefont {C.~C.}\ \bibnamefont
  {Lo}}, \bibinfo {author} {\bibfnamefont {M.}~\bibnamefont {Urdampilleta}},
  \bibinfo {author} {\bibfnamefont {P.}~\bibnamefont {Ross}}, \bibinfo {author}
  {\bibfnamefont {M.~F.}\ \bibnamefont {Gonzalez-Zalba}}, \bibinfo {author}
  {\bibfnamefont {J.}~\bibnamefont {Mansir}}, \bibinfo {author} {\bibfnamefont
  {S.~A.}\ \bibnamefont {Lyon}}, \bibinfo {author} {\bibfnamefont {M.~L.~W.}\
  \bibnamefont {Thewalt}}, \ and\ \bibinfo {author} {\bibfnamefont {J.~J.~L.}\
  \bibnamefont {Morton}},\ }\href {\doibase 10.1038/nmat4250} {\bibfield
  {journal} {\bibinfo  {journal} {Nat. Mater.}\ }\textbf {\bibinfo {volume}
  {14}},\ \bibinfo {pages} {490} (\bibinfo {year} {2015})}\BibitemShut
  {NoStop}%
\bibitem [{\citenamefont {Feher}(1959{\natexlab{a}})}]{feher_electron_1959-1}%
  \BibitemOpen
  \bibfield  {author} {\bibinfo {author} {\bibfnamefont {G.}~\bibnamefont
  {Feher}},\ }\href {\doibase 10.1103/PhysRev.114.1219} {\bibfield  {journal}
  {\bibinfo  {journal} {Phys. Rev.}\ }\textbf {\bibinfo {volume} {114}},\
  \bibinfo {pages} {1219} (\bibinfo {year} {1959}{\natexlab{a}})}\BibitemShut
  {NoStop}%
\bibitem [{\citenamefont {Breiland}\ \emph {et~al.}(1973)\citenamefont
  {Breiland}, \citenamefont {Harris},\ and\ \citenamefont
  {Pines}}]{breiland_optically_1973}%
  \BibitemOpen
  \bibfield  {author} {\bibinfo {author} {\bibfnamefont {W.~G.}\ \bibnamefont
  {Breiland}}, \bibinfo {author} {\bibfnamefont {C.~B.}\ \bibnamefont
  {Harris}}, \ and\ \bibinfo {author} {\bibfnamefont {A.}~\bibnamefont
  {Pines}},\ }\href {\doibase 10.1103/PhysRevLett.30.158} {\bibfield  {journal}
  {\bibinfo  {journal} {Phys. Rev. Lett.}\ }\textbf {\bibinfo {volume} {30}},\
  \bibinfo {pages} {158} (\bibinfo {year} {1973})}\BibitemShut {NoStop}%
\bibitem [{\citenamefont {Huebl}\ \emph {et~al.}(2008)\citenamefont {Huebl},
  \citenamefont {Hoehne}, \citenamefont {Grolik}, \citenamefont {Stegner},
  \citenamefont {Stutzmann},\ and\ \citenamefont {Brandt}}]{huebl_spin_2008}%
  \BibitemOpen
  \bibfield  {author} {\bibinfo {author} {\bibfnamefont {H.}~\bibnamefont
  {Huebl}}, \bibinfo {author} {\bibfnamefont {F.}~\bibnamefont {Hoehne}},
  \bibinfo {author} {\bibfnamefont {B.}~\bibnamefont {Grolik}}, \bibinfo
  {author} {\bibfnamefont {A.~R.}\ \bibnamefont {Stegner}}, \bibinfo {author}
  {\bibfnamefont {M.}~\bibnamefont {Stutzmann}}, \ and\ \bibinfo {author}
  {\bibfnamefont {M.~S.}\ \bibnamefont {Brandt}},\ }\href {\doibase
  10.1103/PhysRevLett.100.177602} {\bibfield  {journal} {\bibinfo  {journal}
  {Phys. Rev. Lett.}\ }\textbf {\bibinfo {volume} {100}},\ \bibinfo {pages}
  {177602} (\bibinfo {year} {2008})}\BibitemShut {NoStop}%
\bibitem [{\citenamefont {Feher}(1959{\natexlab{b}})}]{feher_donor_1959}%
  \BibitemOpen
  \bibfield  {author} {\bibinfo {author} {\bibfnamefont {G.}~\bibnamefont
  {Feher}},\ }\href {\doibase 10.1016/0022-3697(59)90396-8} {\bibfield
  {journal} {\bibinfo  {journal} {J. Phys. Chem. Solids}\ }\textbf {\bibinfo
  {volume} {8}},\ \bibinfo {pages} {486} (\bibinfo {year}
  {1959}{\natexlab{b}})}\BibitemShut {NoStop}%
\bibitem [{\citenamefont {Hale}\ and\ \citenamefont
  {Mieher}(1969)}]{hale_shallow_1969}%
  \BibitemOpen
  \bibfield  {author} {\bibinfo {author} {\bibfnamefont {E.~B.}\ \bibnamefont
  {Hale}}\ and\ \bibinfo {author} {\bibfnamefont {R.~L.}\ \bibnamefont
  {Mieher}},\ }\href {\doibase 10.1103/PhysRev.184.751} {\bibfield  {journal}
  {\bibinfo  {journal} {Phys. Rev.}\ }\textbf {\bibinfo {volume} {184}},\
  \bibinfo {pages} {751} (\bibinfo {year} {1969})}\BibitemShut {NoStop}%
\bibitem [{\citenamefont {Tyryshkin}\ \emph {et~al.}(2006)\citenamefont
  {Tyryshkin}, \citenamefont {Morton}, \citenamefont {Ardavan},\ and\
  \citenamefont {Lyon}}]{tyryshkin_davies_2006}%
  \BibitemOpen
  \bibfield  {author} {\bibinfo {author} {\bibfnamefont {A.~M.}\ \bibnamefont
  {Tyryshkin}}, \bibinfo {author} {\bibfnamefont {J.~J.~L.}\ \bibnamefont
  {Morton}}, \bibinfo {author} {\bibfnamefont {A.}~\bibnamefont {Ardavan}}, \
  and\ \bibinfo {author} {\bibfnamefont {S.~A.}\ \bibnamefont {Lyon}},\ }\href
  {\doibase doi:10.1063/1.2204915} {\bibfield  {journal} {\bibinfo  {journal}
  {J. Chem. Phys.}\ }\textbf {\bibinfo {volume} {124}},\ \bibinfo {pages}
  {234508} (\bibinfo {year} {2006})}\BibitemShut {NoStop}%
\bibitem [{\citenamefont {Morton}\ \emph {et~al.}(2008)\citenamefont {Morton},
  \citenamefont {Tyryshkin}, \citenamefont {Brown}, \citenamefont {Shankar},
  \citenamefont {Lovett}, \citenamefont {Ardavan}, \citenamefont {Schenkel},
  \citenamefont {Haller}, \citenamefont {Ager},\ and\ \citenamefont
  {Lyon}}]{morton_solid-state_2008}%
  \BibitemOpen
  \bibfield  {author} {\bibinfo {author} {\bibfnamefont {J.~J.~L.}\
  \bibnamefont {Morton}}, \bibinfo {author} {\bibfnamefont {A.~M.}\
  \bibnamefont {Tyryshkin}}, \bibinfo {author} {\bibfnamefont {R.~M.}\
  \bibnamefont {Brown}}, \bibinfo {author} {\bibfnamefont {S.}~\bibnamefont
  {Shankar}}, \bibinfo {author} {\bibfnamefont {B.~W.}\ \bibnamefont {Lovett}},
  \bibinfo {author} {\bibfnamefont {A.}~\bibnamefont {Ardavan}}, \bibinfo
  {author} {\bibfnamefont {T.}~\bibnamefont {Schenkel}}, \bibinfo {author}
  {\bibfnamefont {E.~E.}\ \bibnamefont {Haller}}, \bibinfo {author}
  {\bibfnamefont {J.~W.}\ \bibnamefont {Ager}}, \ and\ \bibinfo {author}
  {\bibfnamefont {S.~A.}\ \bibnamefont {Lyon}},\ }\href {\doibase
  10.1038/nature07295} {\bibfield  {journal} {\bibinfo  {journal} {Nature}\
  }\textbf {\bibinfo {volume} {455}},\ \bibinfo {pages} {1085} (\bibinfo {year}
  {2008})}\BibitemShut {NoStop}%
\bibitem [{\citenamefont {George}\ \emph {et~al.}(2010)\citenamefont {George},
  \citenamefont {Witzel}, \citenamefont {Riemann}, \citenamefont {Abrosimov},
  \citenamefont {Notzel}, \citenamefont {Thewalt},\ and\ \citenamefont
  {Morton}}]{george_electron_2010}%
  \BibitemOpen
  \bibfield  {author} {\bibinfo {author} {\bibfnamefont {R.~E.}\ \bibnamefont
  {George}}, \bibinfo {author} {\bibfnamefont {W.}~\bibnamefont {Witzel}},
  \bibinfo {author} {\bibfnamefont {H.}~\bibnamefont {Riemann}}, \bibinfo
  {author} {\bibfnamefont {N.~V.}\ \bibnamefont {Abrosimov}}, \bibinfo {author}
  {\bibfnamefont {N.}~\bibnamefont {Notzel}}, \bibinfo {author} {\bibfnamefont
  {M.~L.~W.}\ \bibnamefont {Thewalt}}, \ and\ \bibinfo {author} {\bibfnamefont
  {J.~J.~L.}\ \bibnamefont {Morton}},\ }\href {\doibase
  10.1103/PhysRevLett.105.067601} {\bibfield  {journal} {\bibinfo  {journal}
  {Phys. Rev. Lett.}\ }\textbf {\bibinfo {volume} {105}},\ \bibinfo {pages}
  {067601} (\bibinfo {year} {2010})}\BibitemShut {NoStop}%
\bibitem [{\citenamefont {Morley}\ \emph {et~al.}(2010)\citenamefont {Morley},
  \citenamefont {Warner}, \citenamefont {Stoneham}, \citenamefont {Greenland},
  \citenamefont {van Tol}, \citenamefont {Kay},\ and\ \citenamefont
  {Aeppli}}]{morley_initialization_2010}%
  \BibitemOpen
  \bibfield  {author} {\bibinfo {author} {\bibfnamefont {G.~W.}\ \bibnamefont
  {Morley}}, \bibinfo {author} {\bibfnamefont {M.}~\bibnamefont {Warner}},
  \bibinfo {author} {\bibfnamefont {A.~M.}\ \bibnamefont {Stoneham}}, \bibinfo
  {author} {\bibfnamefont {P.~T.}\ \bibnamefont {Greenland}}, \bibinfo {author}
  {\bibfnamefont {J.}~\bibnamefont {van Tol}}, \bibinfo {author} {\bibfnamefont
  {C.~W.~M.}\ \bibnamefont {Kay}}, \ and\ \bibinfo {author} {\bibfnamefont
  {G.}~\bibnamefont {Aeppli}},\ }\href {\doibase 10.1038/nmat2828} {\bibfield
  {journal} {\bibinfo  {journal} {Nat. Mater.}\ }\textbf {\bibinfo {volume}
  {9}},\ \bibinfo {pages} {725} (\bibinfo {year} {2010})}\BibitemShut {NoStop}%
\bibitem [{\citenamefont {Simmons}\ \emph {et~al.}(2011)\citenamefont
  {Simmons}, \citenamefont {Brown}, \citenamefont {Riemann}, \citenamefont
  {Abrosimov}, \citenamefont {Becker}, \citenamefont {Pohl}, \citenamefont
  {Thewalt}, \citenamefont {Itoh},\ and\ \citenamefont
  {Morton}}]{simmons_entanglement_2011}%
  \BibitemOpen
  \bibfield  {author} {\bibinfo {author} {\bibfnamefont {S.}~\bibnamefont
  {Simmons}}, \bibinfo {author} {\bibfnamefont {R.~M.}\ \bibnamefont {Brown}},
  \bibinfo {author} {\bibfnamefont {H.}~\bibnamefont {Riemann}}, \bibinfo
  {author} {\bibfnamefont {N.~V.}\ \bibnamefont {Abrosimov}}, \bibinfo {author}
  {\bibfnamefont {P.}~\bibnamefont {Becker}}, \bibinfo {author} {\bibfnamefont
  {H.-J.}\ \bibnamefont {Pohl}}, \bibinfo {author} {\bibfnamefont {M.~L.~W.}\
  \bibnamefont {Thewalt}}, \bibinfo {author} {\bibfnamefont {K.~M.}\
  \bibnamefont {Itoh}}, \ and\ \bibinfo {author} {\bibfnamefont {J.~J.~L.}\
  \bibnamefont {Morton}},\ }\href {\doibase 10.1038/nature09696} {\bibfield
  {journal} {\bibinfo  {journal} {Nature}\ }\textbf {\bibinfo {volume} {470}},\
  \bibinfo {pages} {69} (\bibinfo {year} {2011})}\BibitemShut {NoStop}%
\bibitem [{\citenamefont {Stich}\ \emph {et~al.}(1996)\citenamefont {Stich},
  \citenamefont {Greulich-Weber},\ and\ \citenamefont
  {Spaeth}}]{stich_electrical_1996}%
  \BibitemOpen
  \bibfield  {author} {\bibinfo {author} {\bibfnamefont {B.}~\bibnamefont
  {Stich}}, \bibinfo {author} {\bibfnamefont {S.}~\bibnamefont
  {Greulich-Weber}}, \ and\ \bibinfo {author} {\bibfnamefont {J.-M.}\
  \bibnamefont {Spaeth}},\ }\href {\doibase 10.1063/1.115726} {\bibfield
  {journal} {\bibinfo  {journal} {Appl. Phys. Lett.}\ }\textbf {\bibinfo
  {volume} {68}},\ \bibinfo {pages} {1102} (\bibinfo {year}
  {1996})}\BibitemShut {NoStop}%
\bibitem [{\citenamefont {Hoehne}\ \emph {et~al.}(2011)\citenamefont {Hoehne},
  \citenamefont {Dreher}, \citenamefont {Huebl}, \citenamefont {Stutzmann},\
  and\ \citenamefont {Brandt}}]{hoehne_electrical_2011}%
  \BibitemOpen
  \bibfield  {author} {\bibinfo {author} {\bibfnamefont {F.}~\bibnamefont
  {Hoehne}}, \bibinfo {author} {\bibfnamefont {L.}~\bibnamefont {Dreher}},
  \bibinfo {author} {\bibfnamefont {H.}~\bibnamefont {Huebl}}, \bibinfo
  {author} {\bibfnamefont {M.}~\bibnamefont {Stutzmann}}, \ and\ \bibinfo
  {author} {\bibfnamefont {M.~S.}\ \bibnamefont {Brandt}},\ }\href {\doibase
  10.1103/PhysRevLett.106.187601} {\bibfield  {journal} {\bibinfo  {journal}
  {Phys. Rev. Lett.}\ }\textbf {\bibinfo {volume} {106}},\ \bibinfo {pages}
  {187601} (\bibinfo {year} {2011})}\BibitemShut {NoStop}%
\bibitem [{\citenamefont {Carr}\ and\ \citenamefont
  {Purcell}(1954)}]{carr_effects_1954}%
  \BibitemOpen
  \bibfield  {author} {\bibinfo {author} {\bibfnamefont {H.~Y.}\ \bibnamefont
  {Carr}}\ and\ \bibinfo {author} {\bibfnamefont {E.~M.}\ \bibnamefont
  {Purcell}},\ }\href {\doibase 10.1103/PhysRev.94.630} {\bibfield  {journal}
  {\bibinfo  {journal} {Phys. Rev.}\ }\textbf {\bibinfo {volume} {94}},\
  \bibinfo {pages} {630} (\bibinfo {year} {1954})}\BibitemShut {NoStop}%
\bibitem [{\citenamefont {Medford}\ \emph {et~al.}(2012)\citenamefont
  {Medford}, \citenamefont {Cywinski}, \citenamefont {Barthel}, \citenamefont
  {Marcus}, \citenamefont {Hanson},\ and\ \citenamefont
  {Gossard}}]{medford_scaling_2012}%
  \BibitemOpen
  \bibfield  {author} {\bibinfo {author} {\bibfnamefont {J.}~\bibnamefont
  {Medford}}, \bibinfo {author} {\bibfnamefont {L.}~\bibnamefont {Cywinski}},
  \bibinfo {author} {\bibfnamefont {C.}~\bibnamefont {Barthel}}, \bibinfo
  {author} {\bibfnamefont {C.~M.}\ \bibnamefont {Marcus}}, \bibinfo {author}
  {\bibfnamefont {M.~P.}\ \bibnamefont {Hanson}}, \ and\ \bibinfo {author}
  {\bibfnamefont {A.~C.}\ \bibnamefont {Gossard}},\ }\href {\doibase
  10.1103/PhysRevLett.108.086802} {\bibfield  {journal} {\bibinfo  {journal}
  {Phys. Rev. Lett.}\ }\textbf {\bibinfo {volume} {108}},\ \bibinfo {pages}
  {086802} (\bibinfo {year} {2012})}\BibitemShut {NoStop}%
\bibitem [{\citenamefont {Castner}(1963)}]{castner_raman_1963}%
  \BibitemOpen
  \bibfield  {author} {\bibinfo {author} {\bibfnamefont {T.~G.}\ \bibnamefont
  {Castner}},\ }\href {\doibase 10.1103/PhysRev.130.58} {\bibfield  {journal}
  {\bibinfo  {journal} {Phys. Rev.}\ }\textbf {\bibinfo {volume} {130}},\
  \bibinfo {pages} {58} (\bibinfo {year} {1963})}\BibitemShut {NoStop}%
\bibitem [{\citenamefont {Hartmann}\ and\ \citenamefont
  {Hahn}(1962)}]{hartmann_nuclear_1962}%
  \BibitemOpen
  \bibfield  {author} {\bibinfo {author} {\bibfnamefont {S.~R.}\ \bibnamefont
  {Hartmann}}\ and\ \bibinfo {author} {\bibfnamefont {E.~L.}\ \bibnamefont
  {Hahn}},\ }\href {\doibase 10.1103/PhysRev.128.2042} {\bibfield  {journal}
  {\bibinfo  {journal} {Phys. Rev.}\ }\textbf {\bibinfo {volume} {128}},\
  \bibinfo {pages} {2042} (\bibinfo {year} {1962})}\BibitemShut {NoStop}%
\bibitem [{\citenamefont {Abragam}\ and\ \citenamefont
  {Goldman}(1978)}]{abragam_principles_1978}%
  \BibitemOpen
  \bibfield  {author} {\bibinfo {author} {\bibfnamefont {A.}~\bibnamefont
  {Abragam}}\ and\ \bibinfo {author} {\bibfnamefont {M.}~\bibnamefont
  {Goldman}},\ }\href {\doibase 10.1088/0034-4885/41/3/002} {\bibfield
  {journal} {\bibinfo  {journal} {Rep. Prog. Phys.}\ }\textbf {\bibinfo
  {volume} {41}},\ \bibinfo {pages} {395} (\bibinfo {year} {1978})}\BibitemShut
  {NoStop}%
\bibitem [{\citenamefont {Can}\ \emph {et~al.}(2015)\citenamefont {Can},
  \citenamefont {Walish}, \citenamefont {Swager},\ and\ \citenamefont
  {Griffin}}]{can_time_2015}%
  \BibitemOpen
  \bibfield  {author} {\bibinfo {author} {\bibfnamefont {T.~V.}\ \bibnamefont
  {Can}}, \bibinfo {author} {\bibfnamefont {J.~J.}\ \bibnamefont {Walish}},
  \bibinfo {author} {\bibfnamefont {T.~M.}\ \bibnamefont {Swager}}, \ and\
  \bibinfo {author} {\bibfnamefont {R.~G.}\ \bibnamefont {Griffin}},\ }\href
  {\doibase 10.1063/1.4927087} {\bibfield  {journal} {\bibinfo  {journal} {J.
  Chem. Phys.}\ }\textbf {\bibinfo {volume} {143}},\ \bibinfo {pages} {054201}
  (\bibinfo {year} {2015})}\BibitemShut {NoStop}%
\end{thebibliography}%
\end{document}